%% file: paperShinjo2024_DTCHeavyHex_v4_arXivVer.tex
\documentclass[
aps,
prx,
twocolumn,
superscriptaddress,
floatfix,
longbibliography,
nofootinbib
]{revtex4-2}
\usepackage{graphicx}
\usepackage{dcolumn}
\usepackage{bm}
\usepackage{ulem} 
\usepackage{amsmath}
\usepackage{amssymb}
\usepackage{txfonts}
\usepackage{hyperref}
\usepackage{color} 
\usepackage{xcolor}
\usepackage{listings}
\usepackage{cprotect}

\usepackage[resetlabels,labeled]{multibib}
\newcites{S}{References}

\graphicspath{{./figures/}}

\newcommand{\im}{\mathrm{i}}

\newcommand{\torino}{{ibm\_torino}}

\hypersetup{
  colorlinks=true,
  linkcolor=[rgb]{0.60,0.00,0.00},
  citecolor=[rgb]{0.00,0.00,0.60},
  urlcolor=[rgb]{0.00,0.00,0.60},
  setpagesize=false
}
\def\avg#1{\hat{#1}_{\rm avg}}

\def\cap#1{{\bf #1}}

\makeatletter
\newsavebox{\@brx}
\newcommand{\vvert}[1][]{\savebox{\@brx}{\(\m@th{#1\vert}\)}%
 \mathclose{\copy\@brx\kern-0.5\wd\@brx\usebox{\@brx}}}
\makeatother

\begin{document}
\title{
Unveiling clean two-dimensional discrete time crystals on a digital quantum computer
}
\author{Kazuya Shinjo}
\affiliation{Computational Quantum Matter Research Team, RIKEN Center for Emergent Matter Science (CEMS), Wako, Saitama 351-0198, Japan}
\author{Kazuhiro Seki}
\affiliation{Quantum Computational Science Research Team, RIKEN Center for Quantum Computing (RQC), Wako, Saitama 351-0198, Japan}
\author{Tomonori Shirakawa}
\affiliation{Quantum Computational Science Research Team, RIKEN Center for Quantum Computing (RQC), Wako, Saitama 351-0198, Japan}
\affiliation{Computational Materials Science Research Team,
RIKEN Center for Computational Science (R-CCS), Kobe, Hyogo 650-0047, Japan}
\affiliation{RIKEN Interdisciplinary Theoretical and Mathematical Sciences Program (iTHEMS), Wako, Saitama 351-0198, Japan}
\affiliation{Computational Condensed Matter Physics Laboratory,
RIKEN Cluster for Pioneering Research (CPR), Saitama 351-0198, Japan}
\author{Rong-Yang Sun}
\affiliation{Quantum Computational Science Research Team, RIKEN Center for Quantum Computing (RQC), Wako, Saitama 351-0198, Japan}
\affiliation{Computational Materials Science Research Team,
RIKEN Center for Computational Science (R-CCS), Kobe, Hyogo 650-0047, Japan}
\affiliation{RIKEN Interdisciplinary Theoretical and Mathematical Sciences Program (iTHEMS), Wako, Saitama 351-0198, Japan}
\author{Seiji Yunoki}
\affiliation{Computational Quantum Matter Research Team, RIKEN Center for Emergent Matter Science (CEMS), Wako, Saitama 351-0198, Japan}
\affiliation{Quantum Computational Science Research Team,
RIKEN Center for Quantum Computing (RQC), Wako, Saitama 351-0198, Japan}
\affiliation{Computational Materials Science Research Team,
RIKEN Center for Computational Science (R-CCS), Kobe, Hyogo 650-0047, Japan}
\affiliation{Computational Condensed Matter Physics Laboratory,
RIKEN Cluster for Pioneering Research (CPR), Saitama 351-0198, Japan}

\date{\today}
 
\begin{abstract}
In periodically driven (Floquet) systems, evolution typically results in an infinite-temperature thermal state due to continuous energy absorption over time. 
However, before reaching thermal equilibrium, such systems may transiently pass through a meta-stable state known as a prethermal state. 
This prethermal state can exhibit phenomena not commonly observed in equilibrium, such as discrete time crystals (DTCs), making it an intriguing platform for exploring out-of-equilibrium dynamics.
Here, we investigate the relaxation dynamics of initially prepared product states under periodic driving in a kicked Ising model using the IBM Quantum Heron processor, comprising 133 superconducting qubits arranged on a heavy-hexagonal lattice, over up to $100$ time steps. 
We identify a clean two-dimensional DTC characterised by magnetisation measurements oscillating at twice the period of the Floquet cycle and demonstrate its robustness against perturbations to the transverse field.
This stability does not rely on many-body localisation or on high-frequency Floquet prethermalisation, but emerges in a clean, disorder-free setting.
Moreover, we discover that the longitudinal field induces additional amplitude modulations in the magnetisation with a period incommensurate with the driving period, leading to the emergence of an incommensurately modulated discrete time-crystal (IM-DTC) response. 
These observations are further validated through comparison with tensor-network and state-vector simulations.
Our findings not only provide insight into clean DTC and IM-DTC dynamics in two dimensions but also highlight the utility of gate-based quantum computers for simulating the dynamics of quantum many-body systems, complementing state-of-the-art classical simulations in regimes where entanglement growth challenges their convergence.
\end{abstract}
\maketitle

%

Periodically driven (Floquet) systems host novel phases of matter inaccessible in thermal equilibrium. Notably, discrete time crystals (DTCs)~\cite{Sacha2018, Khemani2019, Else2020, Guo2020, Sacha2020, Zaletel2023} represent genuine out-of-equilibrium phases of matter~\cite{Wilczek2012,Bruno2013Letter, Watanabe2015} feasible in Floquet systems~\cite{Sacha2015, Else2016, Khemani2016}. 
A DTC is characterised by subharmonic responses breaking discrete time-translational symmetry imposed by the periodic drive. 
However, sustaining DTCs as transient meta-stable states faces challenges due to thermalisation, where many-body interactions drive low-entangled states to highly entangled, high-energy states. Overcoming this obstacle requires imparting a many-body localised nature to the dynamics.

One strategy to circumvent rapid thermalisation in driven systems is by introducing disorder in the Floquet Hamiltonian, inducing many-body localisation (MBL) to break ergodicity~\cite{Else2016, Khemani2016, vonKeyserlingk2016, Yao2017, Zhang2017}. 
Recently, disorder-induced MBL-based DTCs (MBL-DTCs) have been demonstrated on digital quantum computers in one dimension~\cite{Ippoliti2021, Mi2022, Frey2022, Zhang2022}.
Furthermore, topological time crystalline order has been achieved in a periodically driven disordered toric code on a superconducting quantum computer~\cite{Wahl2021,Xiang2024}. 
Another avenue for DTCs involves the prethermal regime of periodically driven clean systems in two or higher dimensions~\cite{Else2017, Huang2018, Pizzi2019, Machado2020, Kyprianidis2021, Pizzi2021, Collura2022, Santini2022, Beatrez2023}.
Unlike MBL-DTCs, prethermal DTCs are not stabilised in one-dimensional systems with short-range interactions, aligning with the absence of symmetry breaking at finite temperatures in one dimension. 
Therefore, realising a clean DTC requires two or higher dimensions, or otherwise long-range interactions.

Another avenue for DTCs involves periodically driven clean systems operated in a Floquet prethermal regime, where a sufficiently large driving frequency gives rise to long-lived time-crystalline responses governed by an effective static Hamiltonian~\cite{Else2017, Machado2020, Kyprianidis2021, Beatrez2023}.
In addition, it has been demonstrated that robust DTC behaviour can also occur in clean, strongly interacting systems at finite driving frequencies, beyond the strict high-frequency limit and without relying on MBL protection~\cite{Huang2018, Pizzi2021, Collura2022, Santini2022}.
In such disorder-free settings, one-dimensional systems tend to thermalise, whereas higher-dimensional lattices or effectively long-range couplings can sustain substantially longer-lived time-crystalline behaviour.
This suggests that realising a clean DTC generically requires either two or higher spatial dimensions, or sufficiently long-range interactions, and our results below provide an explicit realisation of the former scenario on a two-dimensional heavy-hexagonal lattice at finite driving frequency, without invoking MBL or high-frequency Floquet prethermalisation.

In simulating dynamics of quantum many-body systems in two dimensions, tensor-network methods have been extensively utilised for large systems beyond the capabilities of state-vector simulations~\cite{Orus2019,Weimer2021,Cirac2021,Tindall23b, Patra2024}.
However, accurate tensor-network simulations over extended periods become challenging in two dimensions due to breakdowns in low-rank tensor approximations when entanglement exceeds certain thresholds dictated by bond dimensions. 
Conversely, recent advancements in noisy intermediate-scale quantum devices have introduced digital quantum computers as another tool to investigate out-of-equilibrium phases of matter, including DTCs. 
Despite these advances, digital quantum simulations of DTCs in genuinely two-dimensional geometries with system sizes comparable to state-of-the-art classical tensor-network simulations remain largely unexplored, and a systematic examination of their stability in such large-scale settings is still lacking.

\begin{figure*}[htbp]
\includegraphics[width=1.0\textwidth]{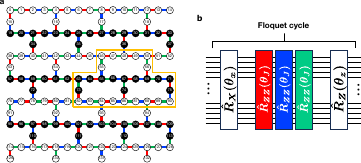}
\caption{
{\bf Two-qubit gate connectivity on a heavy-hexagonal lattice and initial product states.}
	\cap{a}, The overall device geometry of {\torino}, comprising a hevey-hexagonal lattice of $L=133$ qubits. Each circle represents a qubit and the edges indicate the qubit connectivity. Three layers of $R_{ZZ}$ gates in a single Floquet circle are highlighted in red, blue, and green (also see \cap{b}).   
    The enclosed area marked by the yellow line represents the system of $L=28$ qubits, utilised for comparison with state-vector simulations.  
  The initial state is prepared as a product state with qubits denoted by white (black) circles initialised to be $|0\rangle$ ($|1\rangle$). 
  \cap{b}, Schematic representation of the single-cycle Floquet operator $\hat{U}_\mathrm{F}$. The red, blue, and green boxes correspond to the three layers of $R_{ZZ}$ gates, each applied in parallel, indicated in \cap{a}, while the white boxes represent the products of $R_X$ and $R_Z$ gates. 
  The horizontal lines represent the qubits on which the quantum gates operate.
  }
\label{fig:geometry}
\end{figure*}

Here, we demonstrate the realisation of clean DTCs on a two-dimensional heavy-hexagonal lattice of $133$ qubits (see Fig.~\ref{fig:geometry}a) using an IBM Quantum Heron processor, \torino. 
By applying periodic driving to initial product states in a kicked Ising model~\cite{Pineda2014}, involving both transverse and longitudinal fields, we measure local magnetisation to observe its subharmonic response.
With a simple error mitigation protocol based on a depolarising noise model, our results are first validated by showing agreement with both tensor-network simulations of the $133$-qubit system and state-vector simulations of a $28$-qubit system for up to $50$ time steps. 
We then observe a subharmonic period-doubling response of local magnetisation persisting for at least $100$ time steps, confirming its stability against perturbations to the transverse field, which thereby provides evidence for a realisation of prethermal DTCs in two dimensions.
Furthermore, we observe other longer-period subharmonic responses with frequencies incommensurate with the driving period, which we identify as incommensurately modulated DTC (IM-DTC) responses~\cite{Pizzi2019, Pizzi2021}.


We explore the Floquet dynamics of a kicked Ising model on an $L$-qubit system governed by a time-dependent Hamiltonian of period $T$, satisfying $\hat{H}(t)=\hat{H}(t+T)$ with
\begin{equation}\label{eq-hamiltonian}
\hat{H}(t)=
\left\{ \,
    \begin{aligned}
    & h_{x} \sum_{i=0}^{L-1}\hat{X}_{i}, &\text{  for  }0\leq t < T/2\\
    & h_{z} \sum_{i=0}^{L-1}\hat{Z}_{i} - J \sum_{\langle i,j\rangle}\hat{Z}_{i}\hat{Z}_{j} &\text{  for  }T/2\leq t < T
    \end{aligned}
\right.,		
\end{equation} 
where $\hat{X}_i$ and $\hat{Z}_i$ are Pauli operators at qubit $i$, and $\sum_{i}$ and $\sum_{\langle i,j\rangle}$ run over all vertices and edges of the lattice, respectively. $h_x$, $h_z$, and $J$ are parameters, referred to as the transverse field, longitudinal field, and exchange interaction, respectively. 
The associated single-cycle Floquet operator $\hat{U}_\text{F}$
can be expressed in terms of single- and two-qubit
gates as 
\begin{align} \label{eq:uf}
\hat{U}_\text{F}=
\left[ \prod_{i} \hat{R}_{Z_{i}}(\theta_{z}) \right]
\left[ \prod_{\langle i,j\rangle} \hat{R}_{Z_{i}Z_{j}}(\theta_{J}) \right]
\left[ \prod_{i} \hat{R}_{X_{i}}(\theta_{x}) \right],
\end{align}
where 
$\hat{R}_{Z_{i}Z_{j}}(\theta_{J})=\exp\left[-\im \theta_{J} \hat{Z}_i \hat{Z}_j/2\right]$, 
$\hat{R}_{Z_{i}}(\theta_{z}) =\exp\left[-\im \theta_{z} \hat{Z}_i/2\right]$, and 
$\hat{R}_{X_{i}}(\theta_{x}) =\exp\left[-\im \theta_{x} \hat{X}_i/2\right]$ are $ZZ$, $Z$, and $X$ rotation gates with rotation angles $\theta_{J}=-JT$, $\theta_{z}=h_{z}T$, and  $\theta_{x}=h_{x}T$, respectively. 
Since each qubit is coupled to at most three adjacent qubits on the heavy-hexagonal lattice, operation of all the two-qubit gates $\hat{R}_{Z_{i}Z_{j}} (\theta_{J})$ has to be divided into three layers (see Fig.~\ref{fig:geometry}b). 
Each layer consists of $\hat{R}_{Z_{i}Z_{j}} (\theta_{J})$ gates on red, blue, or green edges in Fig.~\ref{fig:geometry}a,
allowing for parallel operation. 
In our implementation on \torino, the Floquet time evolution over $n$ periods is realised as a gate-based (digital) quantum circuit obtained by repeating this sequence of single- and two-qubit gates $n$ times.

The time-evolved state at stroboscopic times $t=nT$ with integer $n$ is expressed as 
$
    |\psi(t)\rangle = (\hat{U}_\mathrm{F})^n |\psi(0)\rangle,
$    
where $|\psi(0)\rangle$ represents the initial state. 
Our primary focus lies in measuring local magnetisation defined as  
$
 \langle \hat{Z}_{j}(t) \rangle=\langle \psi(t)|\hat{Z}_{j}|\psi(t)\rangle,
$
where $\hat{Z}_{j}(t)=(\hat{U}_\mathrm{F}^\dag)^n \hat{Z}_{j} (\hat{U}_\mathrm{F})^n$ is the Heisenberg representation of the Pauli $\hat{Z}_{j}$ operator, and $\langle \cdots \rangle = \langle \psi(0)|\cdots |\psi(0)\rangle$ denotes the expectation value with respect to the initial state.
The initial state is prepared as a product state in the computational basis, forming a stripe pattern of $|0\rangle$'s and $|1\rangle$'s, represented by white and black circles in Fig.~\ref{fig:geometry}a.
Among the three independent model parameters, we set $\theta_{J}=-\pi/2$ and vary the other two parameters $\theta_x$ and $\theta_z$. 
The gate $\hat{R}_{Z_i Z_j}(\theta_J)$ at $\theta_{J}=-\pi/2$ 
is decomposed into the CZ gate, the native two-qubit gate of \torino, and the $S$ gate, as $\hat{R}_{Z_i Z_j}(-\pi/2)=e^{i\pi/4}\widehat{\mathrm{CZ}}_{ij} (\hat{S}_i^\dag \otimes \hat{S}_j^\dag)$. 

For convenience, we introduce a perturbation parameter $\epsilon$ to the transverse field as
$
2 \epsilon = \pi - \theta_x.
$
When $\epsilon=0$, the dynamics of $\hat{Z}_j (t)$ becomes trivial because a single Floquet cycle simply flips the sign of the local magnetisation, $\hat{Z}_j(t+T) = \hat{U}_\mathrm{F}^\dag \hat{Z}_{j}(t) \hat{U}_\mathrm{F} =(-1) \hat{Z}_{j}(t)$. 
This demonstrates that a period-doubling DTC with $|\langle \hat{Z}_j(t) \rangle| = 1$ is realised, at least at the fine-tuned parameter $\theta_{x}=\pi$.
Our primary interest therefore lies in the subharmonic response of the magnetisation for $\epsilon>0$.

We utilise \torino, the IBM Quantum Heron gate-based superconducting quantum processor comprising $133$ superconducting qubits arranged on the heavy-hexagonal lattice (see Fig.~\ref{fig:geometry}a)~\cite{McKay2023}. 
The median infidelity of the native two-qubit gates (i.e., CZ gates) is approximately $4\times 10^{-3}$, while the infidelity of single-qubit gates is around $4 \times 10^{-4}$ with the median read-out error of $1.7 \times 10^{-2}$ (see also Supplementary Information~S1).
Each Floquet cycle involves $150$ CZ gates for the $L=133$ system, totaling $150n$ CZ gates for $n$ time steps. 
Given the three non-parallelisable layers of two-qubit gates per cycle, the circuit depth for the $n$ time steps is $3n$. Considering a two-qubit gate time of about $100$ns, the real-time duration from state preparation to final measurement for the maximum circuit depth $300$ for $n=100$ is estimated as roughly $30 \mu$s, significantly shorter than the median single-qubit relaxation time $T_1=165\mu$s and dephasing time $T_2=138\mu$s.

\begin{figure*}[t]
\includegraphics[width=1.0\textwidth]{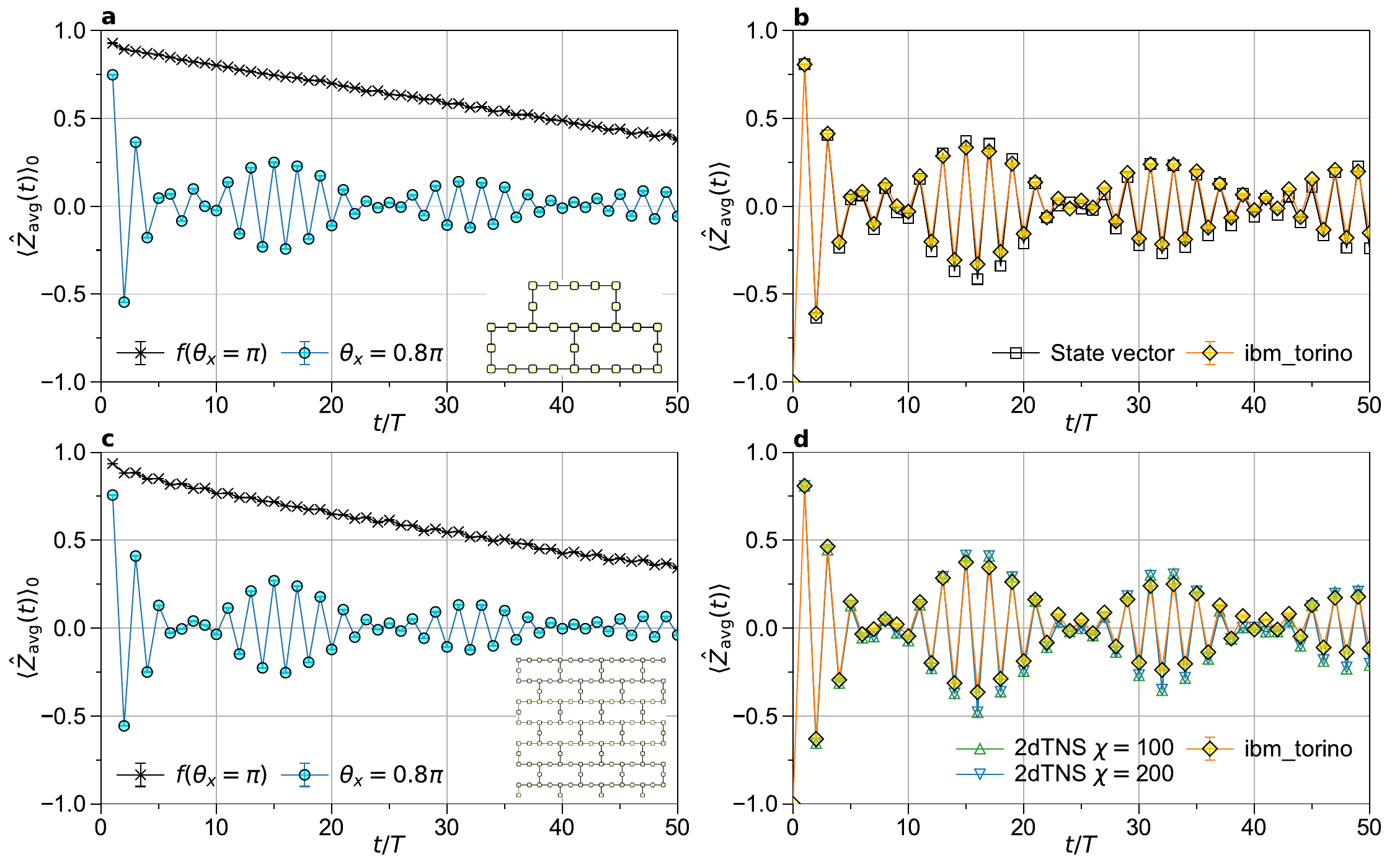}
\caption{
{\bf Error-mitigation protocol and comparison with classical simulations}.
    \cap{a}, Raw data of the averaged magnetisation 
    $\langle \avg{Z}(t)\rangle_{0}$ at $(\theta_x,\theta_z)=(0.8\pi,0.5\pi)$ (cyan circles) and 
    $f(\theta_x=\pi)=|\langle \avg{Z}(t)\rangle_{0,\theta_x=\pi}|$ at $(\theta_x,\theta_z)=(\pi,0.5\pi)$ (black crosses) for $L=28$.
    \cap{b}, Error-mitigated data (yellow diamonds) of the averaged magnetisation $\langle \avg{Z}(t)\rangle$ at $(\theta_x,\theta_z)=(0.8\pi,0.5\pi)$ for $L=28$. 
    Numerically exact results obtained by state-vector simulations are also shown with black squares in \cap{b}.
    \cap{c-d}, Same as \cap{a-b} but for $L=133$.
    Results of classical 2dTNS simulations with bond dimensions $\chi=100$ and 200 are also shown with green and blue triangles, respectively, in \cap{d}. 
    Error bars in \cap{b} and \cap{d} represent the propagated error due to sampling errors in the quantities in the numerator and the denominator of Eq.~(\ref{eq-norm}).
    }
\label{fig:comp}
\end{figure*}

First, we introduce a simple baseline error-mitigation scheme, based on a global depolarising-noise model, to validate that the quantum device provides reliable results for magnetisation dynamics.
We measure the time evolution of the averaged magnetisation over a set of qubits,
\begin{align}\label{eq:Zav}
{\avg{Z}}(t) = \frac{1}{|A|} \sum_{j \in A}   \hat{Z}_{j}(t),   
\end{align}
where $|A|$ denotes the number of qubits in set $A$, and we choose that   
$A=\{ 63, 64, 65, 66, 67, 68, 69, 70, 71 \}$ for the $L=28$ system and 
$A=\{ 57, 58, 59, 60, 61, 62, 63, 64, 65, 66, 67, 68, 69, 70, 71 \}$
for the $L=133$ system (see Fig.~\ref{fig:geometry}a).
In utilising the quantum device, we estimate the expectation value of ${\avg{Z}}(t)$ at each time step by computing the sample mean of outcomes from projective measurements on all qubits within $A$ in the computational basis over $2^{14}$ samples. 
The statistical 
error associated with this estimate is determined as the sample standard deviation 
of the mean. 
The results are shown in Figs.~\ref{fig:comp}a and \ref{fig:comp}c, where neither error-suppression methods such as dynamical decoupling~\cite{Viola1999,Souza2012} nor error-mitigation methods such as zero-noise extrapolation~\cite{Temme2017,Li2017}  and probabilistic error cancellation~\cite{van_den_Berg2023} are used (the same holds for the other results presented below).

As described above, at $\theta_{x}=\pi$, the noiseless expectation value satisfies $|\langle \avg{Z}(t) \rangle|=1$.
However, the absolute values of the raw data obtained from the quantum device, denoted as $f(\theta_x=\pi):=|\langle \avg{Z}(t) \rangle_0|$, deviate from the ideal value $1$, with the deviation increasing over time steps, as observed in Figs.~\ref{fig:comp}a and \ref{fig:comp}c.
To account for this signal decay, we introduce a global depolarising noise model, where the expectation value $\langle \hat{O}(t) \rangle_0$ of an observable $\hat{O}$ subject to depolarising noise is given by 
$
\langle \hat{O}(t) \rangle_0 = 
f \langle \hat{O}(t) \rangle + (1-f) \mathrm{Tr}[\hat{O}(t)]/2^L$~\cite{Swingle2018,Vovrosh2021}.
Here, $f$ is a parameter that characterises the depolarising noise model, with $\langle \hat{O}(t) \rangle $ representing the ideal expectation value of $\hat{O}$ and $\mathrm{Tr}[\hat{O}(t)]/2^L$ being the expectation value over the maximally mixed state. 
Generally, $f$ depends on both the circuit and observable, i.e., $f=f(\theta_J,\theta_x,\theta_z,n,\hat{O})$.  
Since $|\langle \avg{Z}(t)\rangle|=1$ at $\theta_x=\pi$ and $\mathrm{Tr}[\avg{Z}(t)]=0$ as $\hat{Z}_j(t)$ is traceless, the parameter $f$ can be estimated in this trivial case as $f(\theta_J,\pi,\theta_z,n,\avg{Z})=|\langle \avg{Z}(t) \rangle_{0,\theta_x=\pi}|$, where 
$\langle \avg{Z}(t) \rangle_{0,\theta_x=\pi}$ is $\langle \avg{Z}(t) \rangle_{0}$ obtained at $\theta_{x}=\pi$.
For general $\theta_x$, it is difficult to estimate $f$ because the ideal expectation value $\langle \avg{Z}(t) \rangle$ is not available. 
To circumvent this issue, we approximate $f(\theta_J,\theta_x,\theta_z,n,\avg{Z})$ by $f(\theta_J,\pi,\theta_z,n,\avg{Z})=f(\theta_x=\pi)$. 
This approximation leads to the following simple baseline error-mitigation scheme:  
\begin{align}\label{eq-norm}
\langle \avg{Z}(t) \rangle 
\approx
\frac{
\langle \avg{Z}(t) \rangle_{0}}
{|\langle \avg{Z}(t) \rangle_{0,\theta_x=\pi}|}.
\end{align}
We give more detail for this error-mitigation protocol in Supplementary Information~S6.
This procedure effectively compensates a global, configuration-independent decay of the signal and does not attempt full noise cancellation.
Similar approaches have been successfully applied previously to correct magnetisation~\cite{Frey2022} and out-of-time-ordered correlators~\cite{Mi2021}.

The raw data $\langle \avg{Z}(t)\rangle_0$ for the $L=28$ qubit system at $(\theta_{x},\theta_{z})=(0.8\pi,0.5\pi)$ are displayed in Fig.~\ref{fig:comp}a.
These data already capture characteristic oscillations up to 50 time steps, also observed in the state-vector simulation (Fig.~\ref{fig:comp}b). 
However, similar to the trivial case at $\theta_x=\pi$, the signal diminishes with increasing time steps compared to the state-vector simulation results. 
Employing the error-mitigation protocol introduced in Eq.~(\ref{eq-norm}) restores the signal reduction, yielding excellent agreement with the state-vector simulation results up to 50 time steps, as shown in Fig.~\ref{fig:comp}b. 
Further comparisons for other parameters over the extended time steps up to 100 are found in Supplementary Information~S2.

The same error mitigation scheme demonstrates excellent performance even for the $L=133$ system, as shown in Fig.~\ref{fig:comp}c and \ref{fig:comp}d.  
In parallel, we employ a two-dimensional tensor-network state (2dTNS)
method as a classical counterpart (Supplementary Information~S9). 
The 2dTNS results presented here converge with respect to the bond dimension $\chi$, which governs the accuracy of the approximation inherent in the 2dTNS method, for time steps up to at least 50 
(see Supplementary Information~S3 for further comparisons with longer time steps and different parameters). 
Once again, the remarkable agreement between the error-mitigate data and the converged tensor-network simulation results confirms the reliability of the quantum device outcomes in the regime where the classical simulations remain well controlled. 
In this sense, gate-based quantum simulations on the device provide a complementary platform to tensor-network methods for exploring DTC dynamics in two dimensions, particularly at intermediate and long times.
We additionally remark that the present Heron device allows reliable access to substantially more Floquet steps than reported in earlier experiments on comparable systems, such as studies using the IBM Eagle processor of 127 qubits~\cite{Kim2023}, where evolution was typically limited to about 20 time steps with multiple error-mitigation methods.

\begin{figure*}[htbp]
\includegraphics[width=1.0\textwidth]{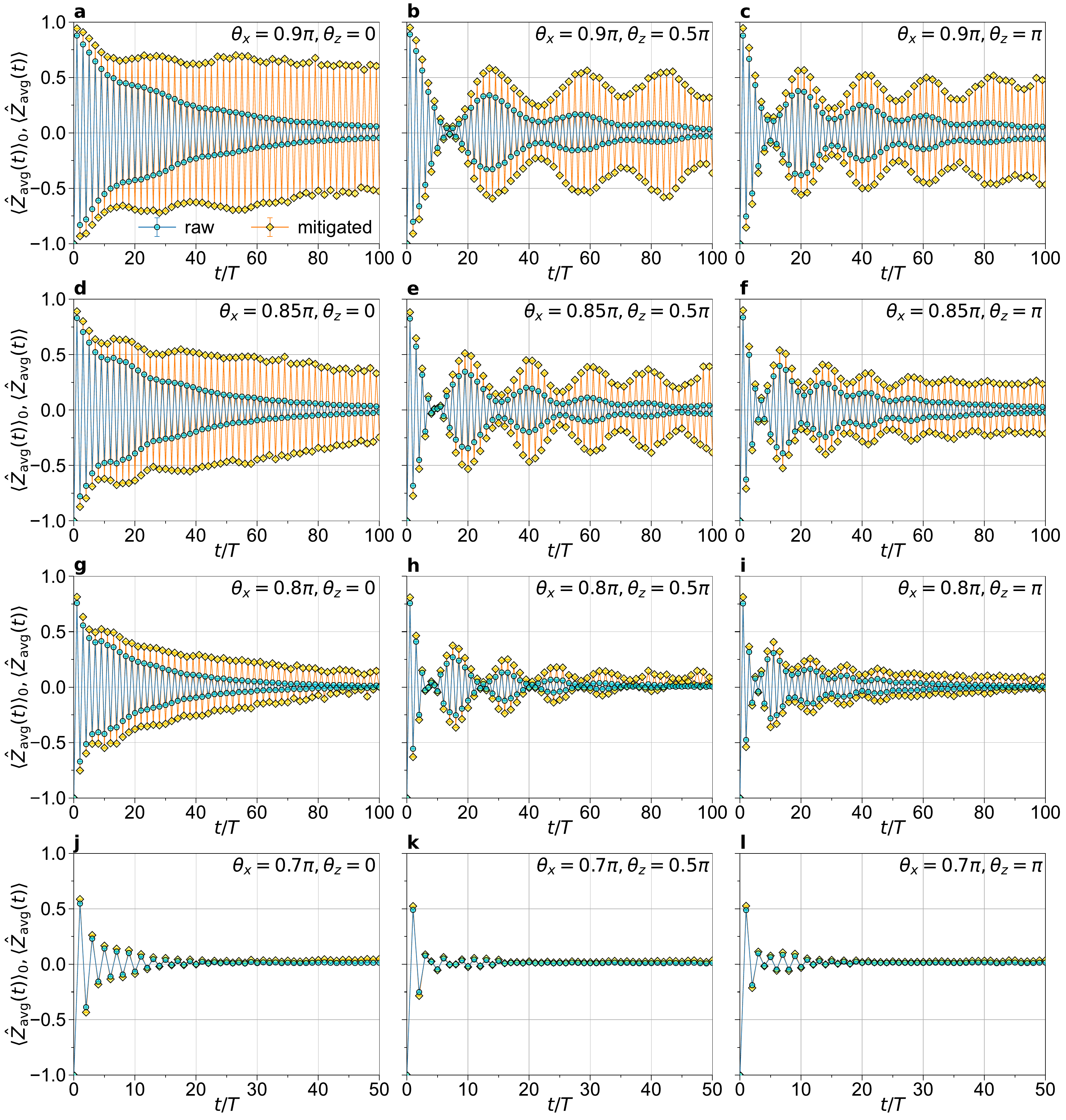}
\caption{
{\bf Dynamics of magnetisation exhibiting DTC, IM-DTC, and thermalisation.}
The raw data $\langle \avg{Z}(t) \rangle_{0}$ (cyan circles) and 
the error-mitigated data $\langle \avg{Z}(t) \rangle$ (yellow diamonds) 
obtained on the heavy-hexagonal lattice of $L=133$ qubits at various parameter sets $(\theta_x,\theta_z)$: 
		\cap{a} $(0.9\pi,0)$, 
		  \cap{b} $(0.9\pi,0.5\pi)$, 
		\cap{c} $(0.9\pi,\pi)$, 
		\cap{d} $(0.85\pi,0)$, 
		\cap{e} $(0.85\pi,0.5\pi)$, 
		\cap{f} $(0.85\pi,\pi)$, 
		\cap{g} $(0.8\pi,0)$, 
		\cap{h} $(0.8\pi,0.5\pi)$, 
		\cap{i} $(0.8\pi,\pi)$, 
		\cap{j} $(0.7\pi,0.0\pi)$,
		\cap{k} $(0.7\pi,0.5\pi)$,
		and \cap{l} $(0.7\pi,\pi)$.
        Notice that the time duration in the horizontal axis in \cap{j}-\cap{l} is half of that in the other panels. 
		}
\label{fig:exper}
\end{figure*}

Having validated the reliability of quantum hardware results, we now delve into discrete time-crystalline orders in two dimensions.  
Figure~\ref{fig:exper} shows the long-time dynamics, spanning up to 100 time steps, of the raw and mitigated magnetisation, $\langle \avg{Z}(t) \rangle_{0}$ and $\langle \avg{Z}(t)  \rangle$, respectively, on the heavy-hexagonal lattice of $L=133$ qubits for various sets of parameters $(\theta_x,\theta_z)$. 
Overall, the decay of the magnetisation becomes more pronounced as $\theta_x$ decreases.  
Specifically, period-doubling oscillations persist even around $t/T=100$ for $\theta_x \geqslant 0.8$, while they are barely observable for $t/T \gtrsim 20$ at $\theta_x = 0.7$, suggesting thermalisation.
We observe similar behavior for the fully-polarised initial state of $|0\rangle$'s (see Supplementary information~S3).
These observations lead to the conclusion that DTCs observed on the heavy-hexagonal lattice remain stable in the range $0.8\pi\leq \theta_{x}\leq \pi$, where a prethermal plateau~\cite{Machado2020} is distinctly visible within the time steps $0 \leq t/T \leq 100$.
This is in sharp contrast to the behavior in one dimension, where magnetisation oscillations quickly decay, irrespective of the parameters $(\theta_x,\theta_z)$ away from the trivial point at $\theta_x=\pi$, as we have also confirmed in Supplementary Information~S4 using the same quantum device.

We emphasise that the clean DTCs on the heavy-hexagonal lattice are quite stable in strongly interacting region $\theta_J=0.5\pi$ (see Supplementary Information~S5).
In Supplementary Information~S7, we confirm that quantum scrambling~\cite{Hayden2007,Sekino2008} has not yet occurred for $\theta_x \leq 0.8\pi$ at least by $t/T<50$.
Since quantum scrambling is an underlying mechanism for the thermalisation of an isolated system~\cite{Deutsch1991,Srednicki1994}, the observed DTCs are stabilised in a prethermal regime.
Compared to a square lattice with the same parameters, the speed of scrambling is much slower in a heavy-hexagonal lattice (see Supplementary Information~S7).
Furthermore, we indeed observe prethermal plateau structures in the magnetisation for small systems (see Supplementary Information~S8), and thus conclude the emergence of prethermal DTCs for a parameter range of $0.8\pi \lesssim \theta_{x} < \pi$.

The clean DTC phase retains its stability even with the addition of longitudinal field $\theta_{z}$, which explicitly breaks Ising symmetry.
Here, the DTC order results from the spontaneous breaking of an emergent Ising symmetry~\cite{vonKeyserlingk2016, Ippoliti2021}.
In addition to the period-doubling oscillation, $\theta_{z}$ induces a longer-period oscillation, as clearly seen in Fig.~\ref{fig:exper} 
(also see Fig.~S6). 

\begin{figure*}[t]
\includegraphics[width=1.0\textwidth]{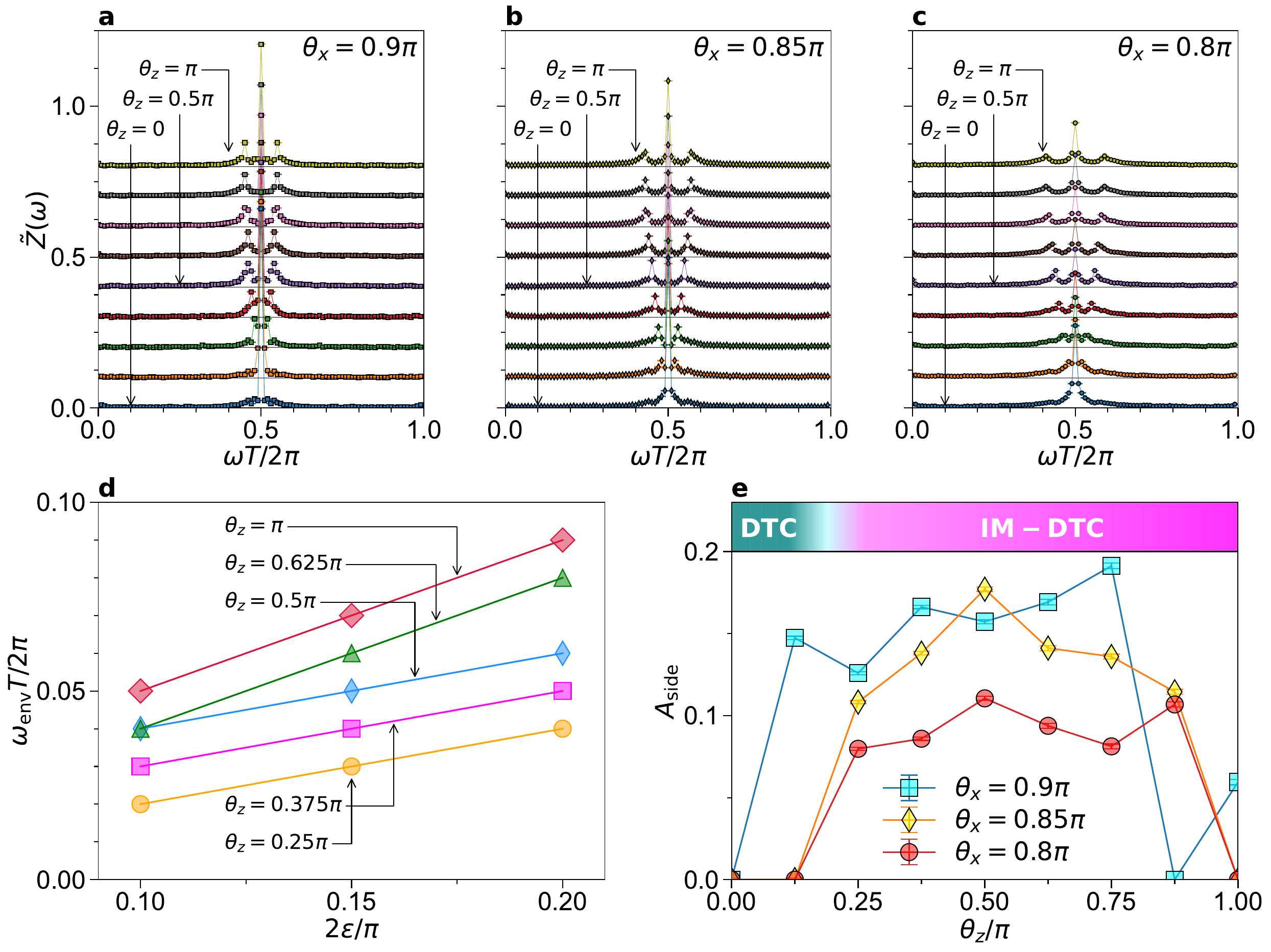}
\caption{{\bf Fourier analysis of incommensurately modulated DTC responses induced by the longitudinal field.} 
Fourier spectrum $\tilde Z(\omega)$ of the averaged magnetisation for 
  \cap{a} $\theta_{x}=0.9\pi$, 
  \cap{b} $\theta_{x}=0.85\pi$, and 
  \cap{c} $\theta_{x}=0.8\pi$ 
  with various values of $\theta_z=0,0.125\pi,0.25\pi,\cdots,\pi$ (from bottom to top). 
  The spectra are vertically offset by 0.05 for clarity.
\cap{d}, Envelope frequency $\omega_\text{env}$ of the IM-DTC modulation as a function of the perturbation parameter $\epsilon$ to the transverse field.
\cap{e}, Intensity of the side peaks, $A_\text{side}$, as a function of $\theta_{z}$. 
A schematic diagram illustrating the crossover between a clean DTC response and an IM-DTC response is shown at the top of the figure.
Error bars in \cap{a}-\cap{c} and \cap{e} represent the propagated error due to sampling errors in $\tilde Z(\omega)$. However, they are smaller than the size of symbols.
		}
\label{fig:FT}
\end{figure*}

To analyse this additional modulation, we perform a discrete Fourier transform
of the error-mitigated magnetisation,
$\tilde Z(\omega)=\bigl|\frac{1}{n_\mathrm{max}}
\sum_{n=0}^{n_\mathrm{max}-1}\langle \avg{Z}(nT) \rangle e^{-i \omega nT }\bigr|$,
where $\omega T=2\pi k/n_\mathrm{max}$ with $k=0,1,\ldots,n_\mathrm{max}-1$, and
$n_\mathrm{max}=100$.  
As shown in Figs.~\ref{fig:FT}a--\ref{fig:FT}c, when $\theta_{z}=0$, the Fourier
spectrum exhibits only a single peak at $\omega T/(2\pi)=0.5$, consistent with a
pure period-doubling DTC response.  
Introducing a longitudinal field produces additional side peaks,
symmetrically located at frequencies $\omega_{\pm}$ around the DTC peak.
These side peaks shift systematically with $\theta_{z}$, reflecting a continuous
and tunable detuning of the modulation frequency.

Although the discrete Fourier frequencies $\omega_{\pm}T/(2\pi)$ are rational
numbers by definition, their smooth dependence on $\theta_{z}$ indicates that,
for fixed parameters and increasing observation time, the corresponding
frequencies fluctuate around generally irrational values.  The appearance of
these additional, parameter-sensitive and essentially incommensurate frequency
components constitutes an IM-DTC response, reminiscent of the behaviour analysed in
Refs.~\cite{Pizzi2019,Pizzi2021}.

The envelope frequency of the longer-period IM-DTC oscillation can be estimated as $\omega_\text{env}=(\omega_{+}-\omega_{-})/2$. 
We observe that the envelope frequency increases proportionally to the perturbation to the transverse field $h_x$, i.e., $\omega_\text{env}\propto \epsilon$ (Fig.~\ref{fig:FT}d).
Particularly at $\theta_{z}=\pi$, $\omega_\text{env} T/(2\pi) = 0.05$, $0.07$, and $0.09$ for $\theta_{x}=0.9\pi$, $0.85\pi$, and $0.8\pi$, respectively.  
These values approximately follow $\omega_\text{env} T/(2\pi)\simeq \epsilon/(2\theta_{J})$,
which is proportional to the frequency of a Bloch oscillation~\cite{Zener1934} induced by a weak transverse field in a one-dimensional Ising model with long-range interactions~\cite{Verdel2020}.
In contrast to the frequency of a period-doubling oscillation, the incommensurate frequency of a side peak $\omega_\pm$ is sensitive to $\epsilon$.
This IM-DTC behavior with modulated period-doubling oscillations induced by a single-color periodic driving leads to the same type of IM-DTC response as proposed in Refs.~\cite{Pizzi2019, Pizzi2021}, but differs from the one in Refs.~\cite{Giergiel2019,Else2020X,He2024}.

To characterise the crossover between a DTC and an IM-DTC response, we plot the sum of the intensities $A_{\pm}$ of the side peaks at $\omega_{\pm}$, denoted as $A_\text{side}=A_{+}+A_{-}$, in Fig.~\ref{fig:FT}e.
$A_\mathrm{side}$ tends to decrease with decreasing $\theta_x$.
At $\theta_{x}=0.7\pi$, $\langle \avg{Z}(t) \rangle$ quickly decays with increasing time step $t/T$ as seen in Figs.~\ref{fig:exper}j-\ref{fig:exper}l, indicating the absence of both a long-lived DTC and an IM-DTC modulation. 
When $\theta_{x}=\pi$, only the period-doubling DTC is present, as $|\langle \avg{Z}(t) \rangle|$ remains 1 regardless of the value of $\theta_{z}$.
Therefore, we conclude that an IM-DTC response manifests in a prethermal regime persisting within the timescale $0 \leq t/T \leq 100$ for a parameter range of $0.8\pi \lesssim \theta_{x} \lesssim 0.9\pi$ and 
$0.25\pi \lesssim \theta_{z} \leq \pi$.


The recent advancements in the quality of quantum devices have enabled the successful simulation of quantum dynamics on a much larger scale, both in terms of qubit count and time step duration. 
Specifically, achieving quantum dynamics simulation on a 133-qubit system for up to 100 time steps using a digital quantum computer, as demonstrated in this study, represents a significant leap forward compared to previous studies. 
Our comparative study, employing a 2dTNS method, confirmed the agreement between the results obtained using the quantum device and those from the classical simulations, for approximately up to 50 time steps. 
However, in our investigation of both 28-qubit and 133-qubit systems, we encountered a critical parameter regime where the perturbation to the transverse field approaches $2\epsilon \approx 0.2\pi$, near the crossover boundary between DTC and IM-DTC responses and thermalisation. 
In this regime, the limitation of classical simulations becomes apparent, particularly for long-time steps around 100. 
Indeed, in the 28-qubit system, we observe discrepancies between the results of the 2dTNS method, even with the largest feasible bond dimension, and those of the state-vector simulation (Supplementary Information~S2). 
Additionally, in the 133-qubit system, a slow convergence of the results of the 2dTNS method with respect to the bond dimension is observed, implying the growth of entanglement that is hardly tractable within the classical resources available for the 2dTNS method (Supplementary Information~S3).
Therefore, our study of Floquet dynamics in this parameter regime, extending up to 100 time steps on the 133-qubit system, pushes the boundaries of classical simulations specially based on tensor-network methods to their limit, emphasising the significant potential of current digital quantum computers for simulating out-of-equilibrium quantum dynamics in two dimensions.
The utilisation of a stabilizer formalism may further improve the simulatability of a tensor network~\cite{Begusic2024,MasotLlima2024}.
It would be a very interesting future challenge to reproduce the DTC and IM-DTC dynamics in the prethermal regime up to 100 time steps using a tensor network method that incorporates quantum noise~\cite{Mangini2024}.

We are grateful for valuable discussions with Netanel Lindner.
This work is based on results obtained in part from a project, JPNP20017, subsidised by the New Energy and Industrial Technology Development Organization (NEDO), Japan. 
We acknowledge the support from the Japan Society for the Promotion of Science (JSPS) KAKENHI Grants (Grant Nos.  
JP19K23433,
JP21H04446, 
JP22K03520, and
JP23K13066) 
from the Ministry of Education, Culture, Sports, Science and Technology (MEXT), Japan. 
We also appreciate the funding received from 
JST COI-NEXT (Grant No. JPMJPF2221) and  
the Program for Promoting Research of the Supercomputer Fugaku (Grant No. MXP1020230411) from MEXT, Japan.  
Furthermore, we acknowledge the support from 
the UTokyo Quantum Initiative, 
the RIKEN TRIP initiative (RIKEN Quantum and Many-body Electron Systems), and
the COE research grant in computational science from Hyogo Prefecture and Kobe City through the Foundation for Computational Science.
Tensor-network simulations are based on the high-performance tensor computing library \textit{GraceQ/tensor}~\cite{gqten}.
A part of the numerical simulations has been performed using the HOKUSAI supercomputer at RIKEN, and the supercomputer system at the D3 center, Osaka University, through the HPCI System Research Project (Project ID: hp250062), and the supercomputer Fugaku installed at RIKEN Center for Computational Science (Project IDs:hp230293 and ra000011).
The code used to generate the 2dTNS simulations is available at \href{https://github.com/r-ccs-cms/2dtn-dtc-2026}{https://github.com/r-ccs-cms/2dtn-dtc-2026}.
\bibliography{bibdtc}

\clearpage

%
%
\renewcommand{\theequation}{S\arabic{equation}}
\setcounter{equation}{0}
\renewcommand\thefigure{S\arabic{figure}}
\setcounter{figure}{0}
\renewcommand\thetable{S\arabic{table}}
\setcounter{table}{0}
\renewcommand{\thesection}{S\arabic{section}}
\renewcommand{\thesubsection}{S\arabic{subsection}}
\setcounter{section}{0}
\renewcommand{\bibnumfmt}[1]{[S#1]}
\renewcommand{\citenumfont}[1]{S#1}

\input{SM_DTC_v4_arXivVer}

\end{document}

%% file: SM_DTC_v4_arXivVer.tex
\onecolumngrid

\begin{center}
{\bf \large Supplementary Information to ``Unveiling clean two-dimensional discrete time crystals on a digital quantum computer''}
\end{center}
\begin{center}
Kazuya Shinjo$^{1}$, Kazuhiro Seki$^{2}$, Tomonori Shirakawa$^{2,3,4,5}$, Rong-Yang Sun$^{2,3,4}$, and Seiji Yunoki$^{1,2,3,5}$\\
\vspace*{0.1cm}
{\footnotesize
$^1$Computational Quantum Matter Research Team, RIKEN Center for Emergent Matter Science (CEMS), Wako, Saitama 351-0198, Japan\\
$^2$Quantum Computational Science Research Team, RIKEN Center for Quantum Computing (RQC), Wako, Saitama 351-0198, Japan\\
$^{3}$Computational Materials Science Research Team, RIKEN Center for Computational Science (R-CCS), Kobe, Hyogo 650-0047, Japan\\
$^4$RIKEN Interdisciplinary Theoretical and Mathematical Sciences Program (iTHEMS), Wako, Saitama 351-0198, Japan\\
$^{5}$Computational Condensed Matter Physics Laboratory, RIKEN Cluster for Pioneering Research (CPR), Saitama 351-0198, Japan
}
\end{center}



\section{Quantum device conditions}\label{sec:device}

The experimental data presented in this study were obtained using the IBM Quantum Heron processor, {\torino}, via cloud access, predominantly during the period from January 1 to January 31, 2024.
Throughout this duration, the average relaxation and dephasing times $T_{1}$ and $T_{2}$, as well as the average readout assignment error rate, across all qubits were 165$\mu$s, 138$\mu$s, and 0.03, respectively. 
Additionally, the average error rate of CZ gates and the average duration of CZ gates across all qubits were 0.006 and 101ns, respectively. 
Notably, no significant deviations were observed in the data obtained on different dates, indicating the consistency and stability of the quantum device utilised for this study.

\section{Results for a heavy-hexagonal lattice of $L=28$ qubits} \label{sec:L28}

\begin{figure*}[t]
\includegraphics[width=1.0\textwidth]{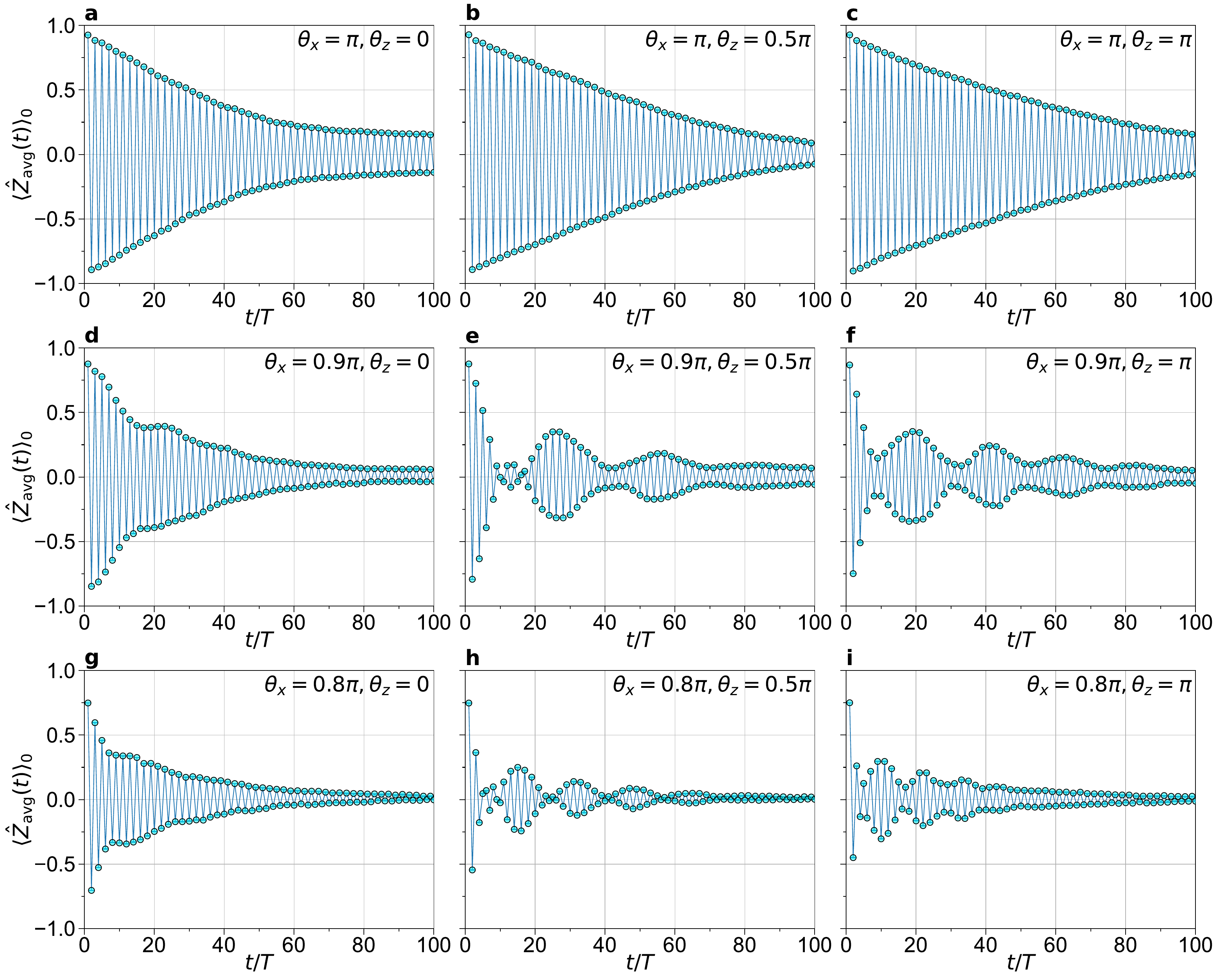}
\caption{Error-unmitigated raw data $\langle \avg{Z}(t) \rangle_{0}$ for the heavy hexagonal lattice of $L=28$ qubits obtained using {\torino} 
(see Fig.~1a). 
The parameters $(\theta_x,\theta_z)$ are 
\cap{a} $(\pi,0)$, 
\cap{b} $(\pi,0.5\pi)$, 
\cap{c} $(\pi,\pi)$, 
\cap{d} $(0.9\pi,0)$, 
\cap{e} $(0.9\pi,0.5\pi)$, 
\cap{f} $(0.9\pi,\pi)$, 
\cap{g} $(0.8\pi,0)$, 
\cap{h} $(0.8\pi,0.5\pi)$, and 
\cap{i} $(0.8\pi,\pi)$. 
Although smaller than the size of symbols in this scale, 
the statistical errors of measurements are estimated in the same manner 
as described in the main text. }
\label{fig:L28_raw}
\end{figure*}

Figure~\ref{fig:L28_raw} shows the error-unmitigated raw data of the averaged magnetisation $\langle \avg{Z}(t) \rangle_{0}$ for up to 100 time steps obtained for the $L=28$ system using {\torino} (see Fig.~1a). Here, the average is performed over 
the same set $A$ of qubits as described in the main text, i.e., 
$A=\{ 63, 64, 65, 66, 67, 68, 69, 70, 71 \}$. 
To mitigate errors in the raw data for $\theta_x\ne\pi$, the results for the trivial cases with $\theta_x=\pi$ are utilised, as detailed in Eq.~(4).

\begin{figure*}[t]
\includegraphics[width=1.0\textwidth]{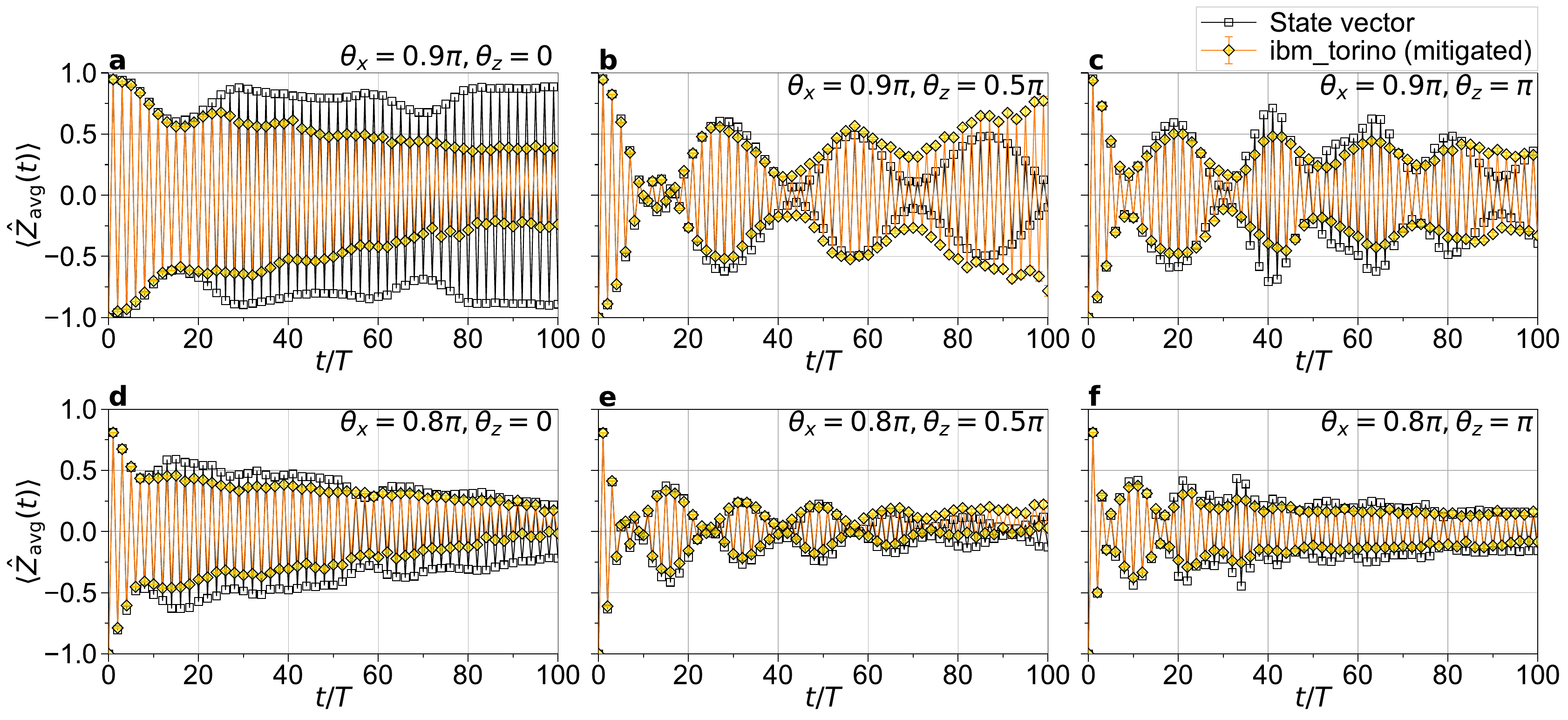}
\caption{
Error-mitigated data $\langle \avg{Z}(t) \rangle$ (yellow diamonds)
are compared with the results obtained by the classical state-vector simulations 
(black squares) for the heavy-hexagonal lattice of $L=28$ qubits. The parameters $(\theta_x,\theta_z)$ are 
\cap{a} $(0.9\pi,0)$, 
\cap{b} $(0.9\pi,0.5\pi)$, 
\cap{c} $(0.9\pi,\pi)$, 
\cap{d} $(0.8\pi,0)$, 
\cap{e} $(0.8\pi,0.5\pi)$, and 
\cap{f} $(0.8\pi,\pi)$. 
The error bars indicated represent the propagated error in the quantities in the numerator and the denominator of Eq.~(4).
}
\label{fig:L28_mit}
\end{figure*}

\begin{figure*}[t]
\includegraphics[width=1.0\textwidth]{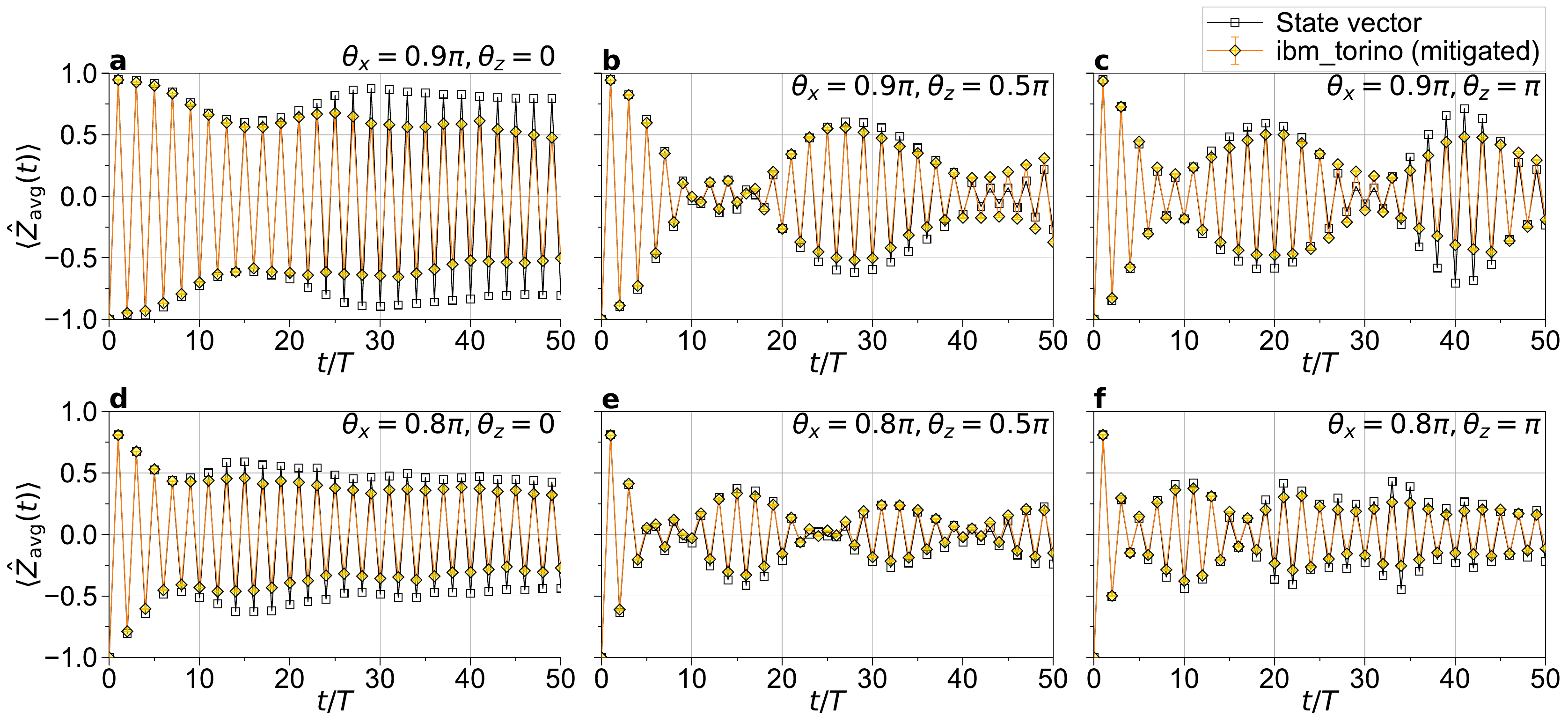}
\caption{
Enlarged plots of Fig.~\ref{fig:L28_mit}, focusing on the time evolution for up to 50 time steps. 
\label{fig:L28_mit_en}
}
\end{figure*}

Comparing with the results obtained by the state-vector method~\cite{S:Seki2022} in Figs.~\ref{fig:L28_mit} and \ref{fig:L28_mit_en}, we find that error-mitigated values $\langle \avg{Z}(t) \rangle$ are generally in good agreement with the numerically exact values, although deviations become apparent in some cases, particularly for long time steps.

\begin{figure*}[t]
\includegraphics[width=1.0\textwidth]{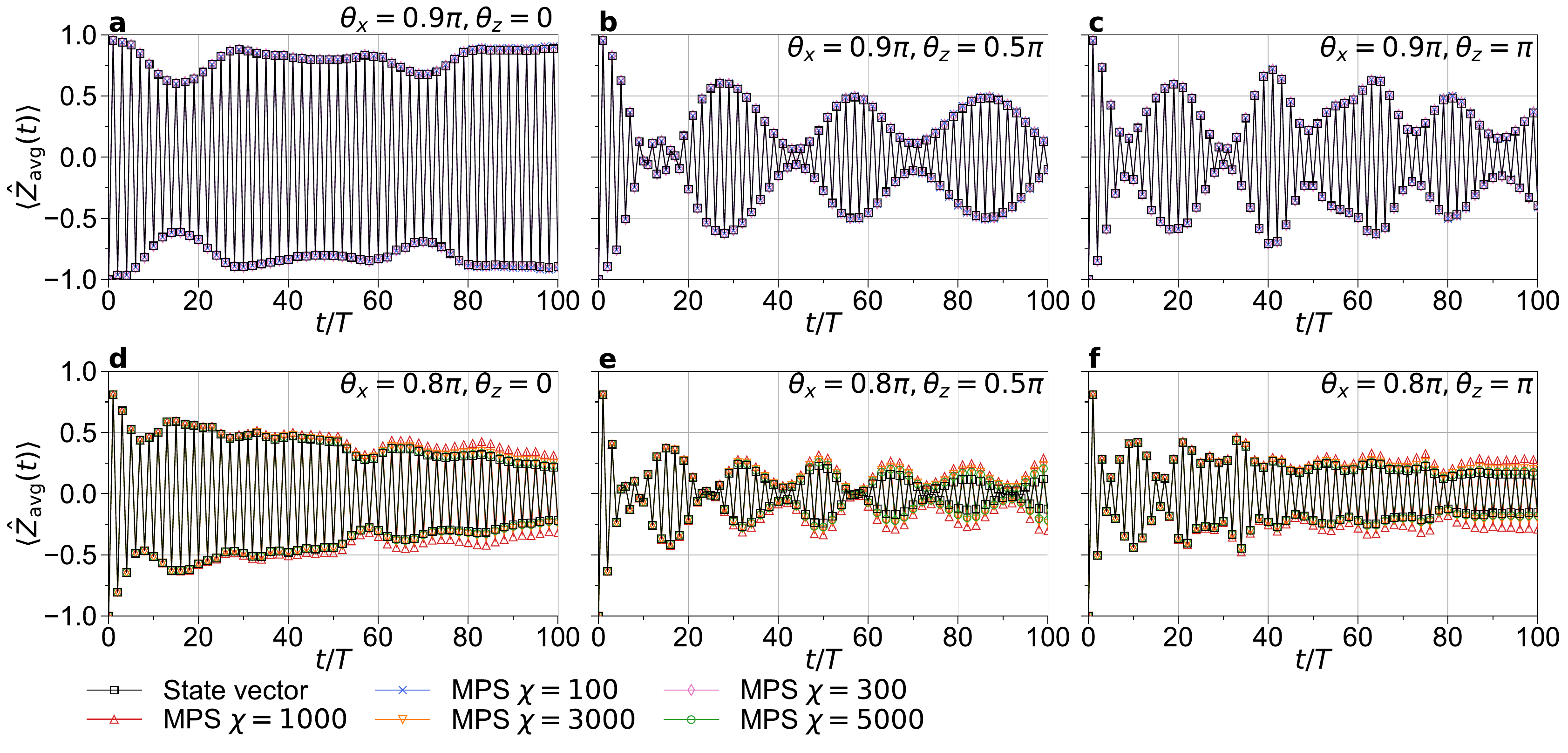}
\caption{
Comparison of the results for $\langle \avg{Z}(t) \rangle$ on the $L=28$ heavy-hexagonal lattice obtained by the MPS method with various bond dimensions ($\chi=100$, 300, 1000, 3000, and 5000) and the state-vector method. 
The parameters $(\theta_x,\theta_z)$ are 
\cap{a} $(0.9\pi,0)$, 
\cap{b} $(0.9\pi,0.5\pi)$, 
\cap{c} $(0.9\pi,\pi)$, 
\cap{d} $(0.8\pi,0)$, 
\cap{e} $(0.8\pi,0.5\pi)$, and 
\cap{f} $(0.8\pi,\pi)$. 
}
\label{fig:L28_mps}
\end{figure*}

\begin{figure*}[t]
\includegraphics[width=1.0\textwidth]{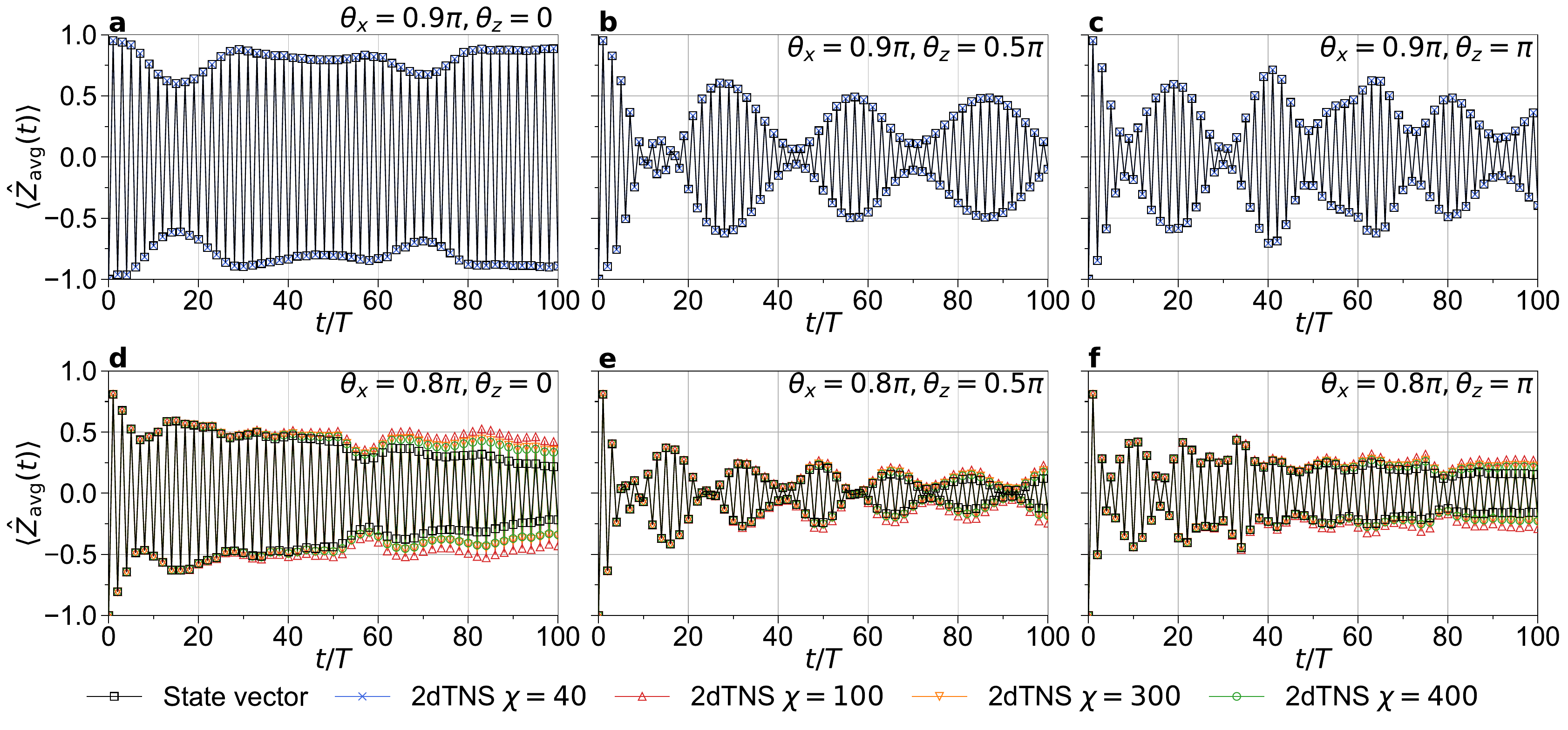}
\caption{
Comparison of the results for $\langle \avg{Z}(t) \rangle$ on the $L=28$ heavy-hexagonal lattice obtained by the 2dTNS method with various bond dimensions ($\chi=40$, 100, 300, and 400) and the state-vector method. 
The parameters $(\theta_x,\theta_z)$ are 
\cap{a} $(0.9\pi,0)$, 
\cap{b} $(0.9\pi,0.5\pi)$, 
\cap{c} $(0.9\pi,\pi)$, 
\cap{d} $(0.8\pi,0)$, 
\cap{e} $(0.8\pi,0.5\pi)$, and 
\cap{f} $(0.8\pi,\pi)$. 
}
\label{fig:L28_peps}
\end{figure*}

To assess the accuracy of various tensor network methods, including 
matrix product state (MPS)~\cite{S:Sun2023} and two-dimensional tensor network state (2dTNS) methods (for details of our 2dTNS simulations, 
see Sec.~\ref{sec:peps}), we compare the results of $\langle \avg{Z}(t) \rangle$ obtained by these methods with the numerically exact results calculated by the state-vector method in Figs.~\ref{fig:L28_mps} and \ref{fig:L28_peps}.
As shown in Figs.~\ref{fig:L28_mps}a-\ref{fig:L28_mps}c and \ref{fig:L28_peps}a-\ref{fig:L28_peps}c, when $\theta_{x}=0.9\pi$, $\langle \avg{Z}(t) \rangle$ obtained by these tensor network methods with relatively small bond dimensions $\chi$ 
(MPS with $\chi=300$ and 2dTNS with $\chi=40$) 
sufficiently converge consistently to the numerically exact values, 
even for time steps up to 100. 
However, as shown in Figs.~\ref{fig:L28_mps}d-\ref{fig:L28_mps}f and \ref{fig:L28_peps}d-\ref{fig:L28_peps}f, when $\theta_{x}=0.8\pi$, which is close to the crossover boundary between the discrete time crystal (DTC) and incommensurately modulated DTC (IM-DTC) responses and the thermalised regime, a larger $\chi$ is required to obtain the converged value, due to the increase of entanglement. 
For example, when $(\theta_x,\theta_z) = (0.8\pi,0)$ 
(Fig.~\ref{fig:L28_peps}d), 2dTNS with $\chi=400$ is insufficient to obtain the numerically exact results for $t\agt 50$. 
A similar trend is observed for other parameters shown in Figs.~\ref{fig:L28_peps}e and \ref{fig:L28_peps}f and for the MPS method shown in Figs.~\ref{fig:L28_mps}d-\ref{fig:L28_mps}f. 
These results suggest that the long-time time-evolution simulations of $\langle \avg{Z}(t)\rangle$ based on classical tensor network methods already face challenges even for $L=28$ in a parameter region around $\theta_{x}=0.8\pi$. 
We expect to encounter similar challenges for $L=133$, where the numerically exact state-vector method cannot be applied, although a recent report suggests that the accuracy of a 2dTNS method improves as the system size increases~\cite{S:Tindall23b}.  
This emphasises the significance of utilising a quantum computer precisely in this parameter regime. 


\section{Results for a heavy-hexagonal lattice of $L=133$ qubits}\label{sec:L133}

\begin{figure*}[t]
\includegraphics[width=1.0\textwidth]{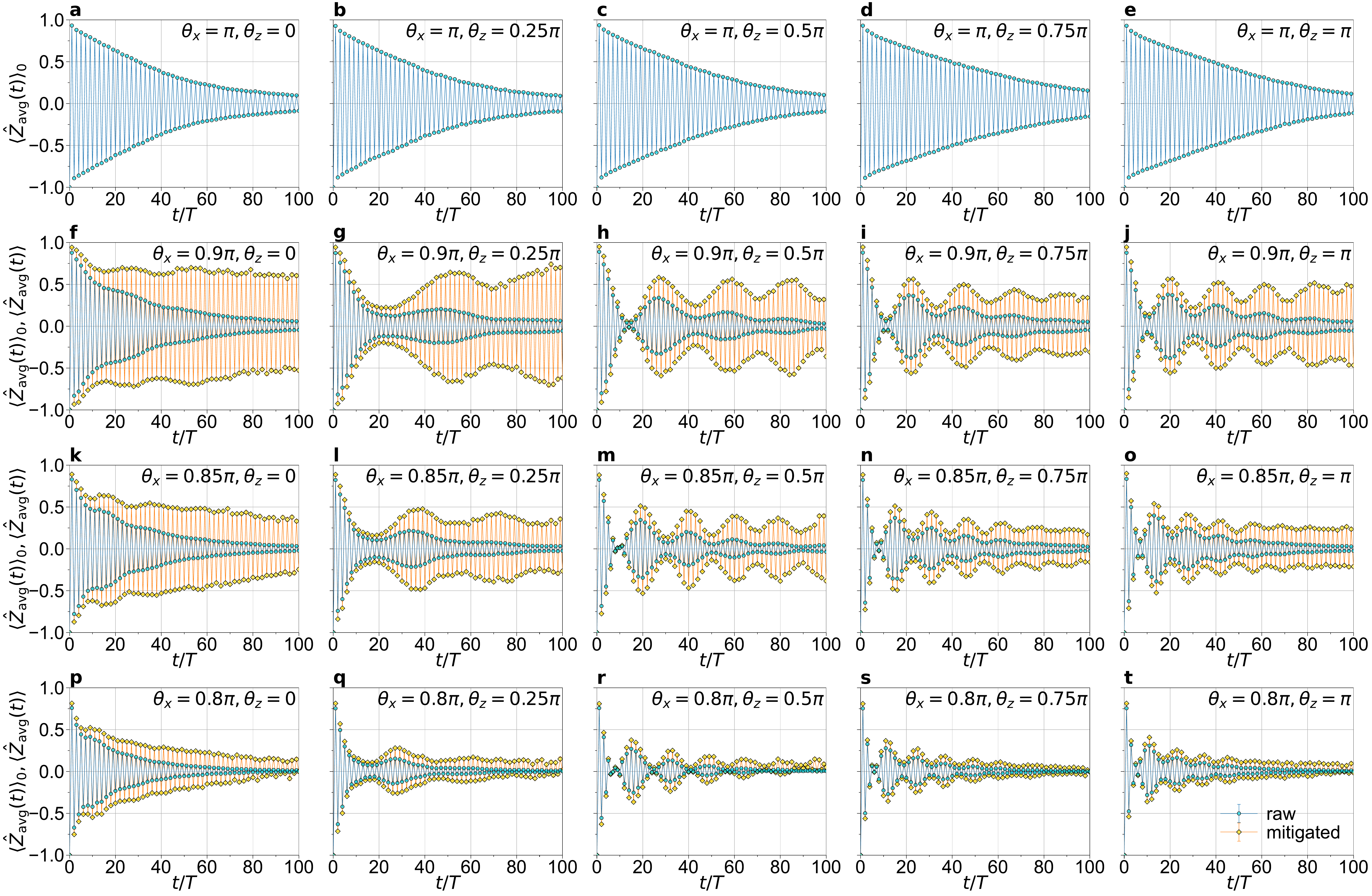}
\caption{
Error-unmitigated raw data $\langle \avg{Z}(t) \rangle_{0}$ (cyan circles) and error-mitigated data $\langle \avg{Z}(t) \rangle$ (yellow diamonds) for the heavy-hexagonal lattice of $L=133$ qubits obtained using {\torino} 
(see Fig.~1a). The parameters 
$(\theta_x,\theta_z)$ are 
\cap{a} $(\pi,0)$, 
\cap{b} $(\pi,0.25\pi)$, 
\cap{c} $(\pi,0.5\pi)$, 
\cap{d} $(\pi,0.75\pi)$, 
\cap{e} $(\pi,\pi)$, 
\cap{f} $(0.9\pi,0)$, 
\cap{g} $(0.9\pi,0.25\pi)$, 
\cap{h} $(0.9\pi,0.5\pi)$, 
\cap{i} $(0.9\pi,0.75\pi)$, 
\cap{j} $(0.9\pi,\pi)$, 
\cap{k} $(0.85\pi,0)$, 
\cap{l} $(0.85\pi,0.25\pi)$, 
\cap{m} $(0.85\pi,0.5\pi)$, 
\cap{n} $(0.85\pi,0.75\pi)$, 
\cap{o} $(0.85\pi,\pi)$, 
\cap{p} $(0.8\pi,0)$, 
\cap{q} $(0.8\pi,0.25\pi)$, 
\cap{r} $(0.8\pi,0.5\pi)$, 
\cap{s} $(0.8\pi,0.75\pi)$, and 
\cap{t} $(0.8\pi,\pi)$. 
The error-mitigated data for the cases of $\theta_x=\pi$ are not shown because they are $\langle \avg{Z}(t) \rangle=\pm 1$ by definition.
		}
\label{fig:L133_raw}
\end{figure*}

Figure~\ref{fig:L133_raw} shows the error-unmitigated raw data of the average magnetisation $\langle \avg{Z}(t) \rangle_{0}$ and the error-mitigated data $\langle \avg{Z}(t) \rangle$ for up to 100 time steps obtained for the $L=133$ system using {\torino} 
(see Fig.~1a).
Here, the average is performed over the same set $A$ of qubits as described in the main text, i.e., 
$A=\{ 57, 58, 59, 60, 61, 62, 63, 64, 65, 66, 67, 68, 69, 70, 71 \}$. 
Remarkably, even without error mitigation, distinct signals of period-doubling oscillations persist for over 100 time steps. 
To mitigate errors in the raw data for $\theta_x\ne\pi$, the results corresponding to the trivial cases with $\theta_x=\pi$ are utilised, 
as explained in Eq.~(4).

For the trivial cases with $\theta_{x}=\pi$, as shown in Figs.~\ref{fig:L133_raw}a-\ref{fig:L133_raw}e, $|\langle \avg{Z}(t) \rangle_{0}|$ should ideally be 1. However, deviations from this ideal value are observed, which become more pronounced with increasing time steps, owing to the noise and decoherence inherent in the quantum device. 
The number of CZ gates in the quantum circuit increases by 150 for each operation of $\hat{U}_\text{F}$ as defined in Eq.~(2).
At $t/T=100$, the quantum circuit volume $v = 15000$ (defined by the total number of CZ gates) superficially implies $|\langle \avg{Z}(t=100T) \rangle_{0}| = (1-p)^{v}\simeq 10^{-27}$, with a typical two-qubit infidelity in {\torino} of $p=4\times 10^{-3}$.
However, in practice, we observe $|\langle \avg{Z}(t=100T) \rangle_{0}|\simeq 0.1$ in Figs.~\ref{fig:L133_raw}a-\ref{fig:L133_raw}e, suggesting an effective quantum circuit volume $v_\text{eff}\simeq 600$. 
Since we measure local observables, $v_\text{eff}$ is significantly smaller than $v$, as discussed in Ref.~\cite{S:Kechedzhi2024}. 
Additionally, it is interesting to notice that the raw data for the $L=28$ system with the trivial parameters, as shown in Figs.~\ref{fig:L28_raw}a-\ref{fig:L28_raw}c, also exhibit $|\langle \avg{Z}(t=100T) \rangle_{0}|\simeq 0.1-0.15$, indicating a similar $v_\text{eff}$.

\begin{figure*}[t]
\includegraphics[width=1.0\textwidth]{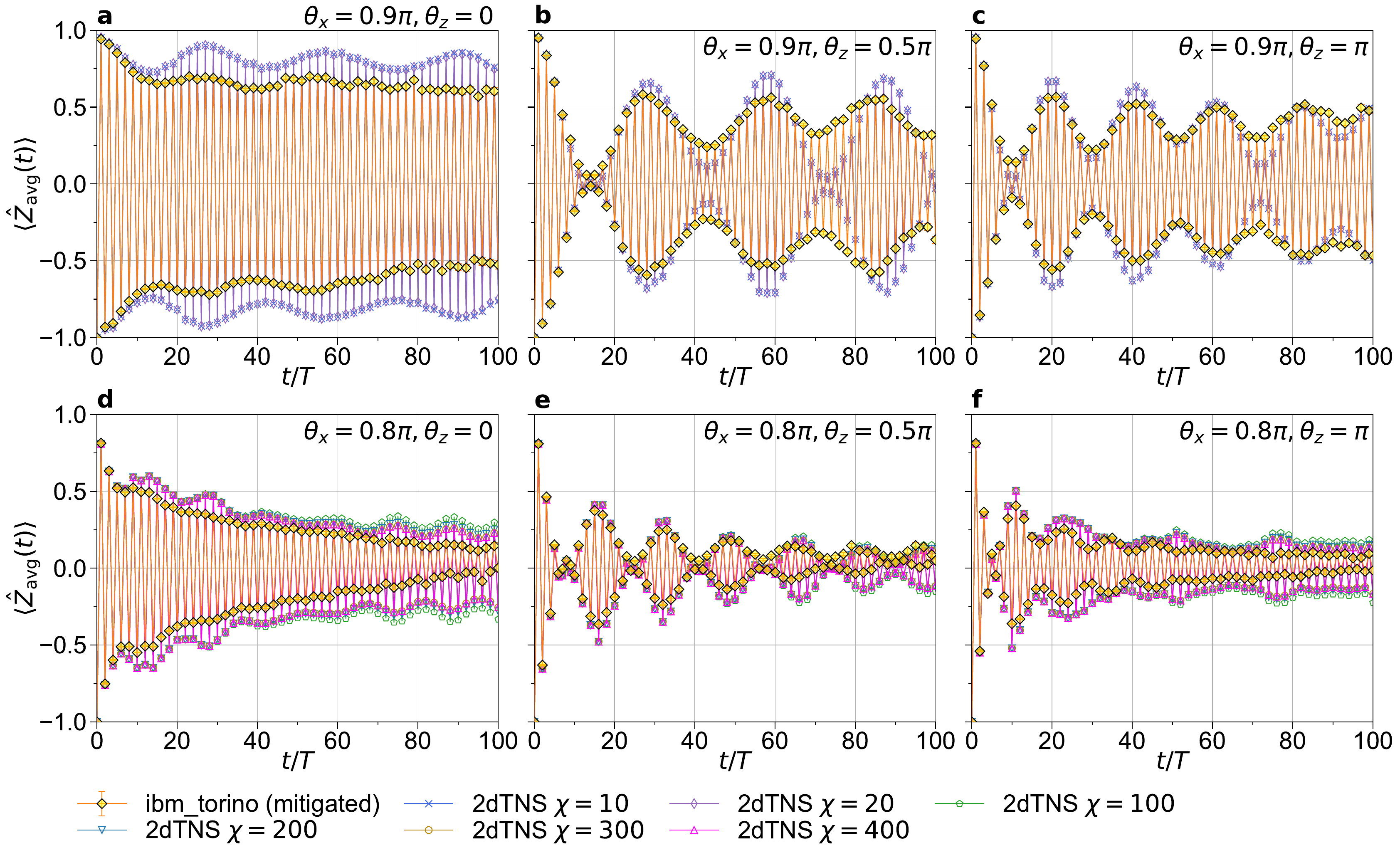}
\caption{
Error-mitigated data $\langle \avg{Z}(t) \rangle$ are compared with the results obtained by the 2dTNS method with various bond dimensions 
($\chi=10$, 20, 100, 200, 300, and 400) for the heavy-hexagonal lattice of 
$L=133$ qubits. 
The parameters $(\theta_x,\theta_z)$ are
\cap{a} $(0.9\pi,0)$, 
\cap{b} $(0.9\pi,0.5\pi)$, 
\cap{c} $(0.9\pi,\pi)$, 
\cap{d} $(0.8\pi,0)$, 
\cap{e} $(0.8\pi,0.5\pi)$, and 
\cap{f} $(0.8\pi,\pi)$. 
}
\label{fig:L133_peps}
\end{figure*}

\begin{figure*}[t]
\includegraphics[width=1.0\textwidth]{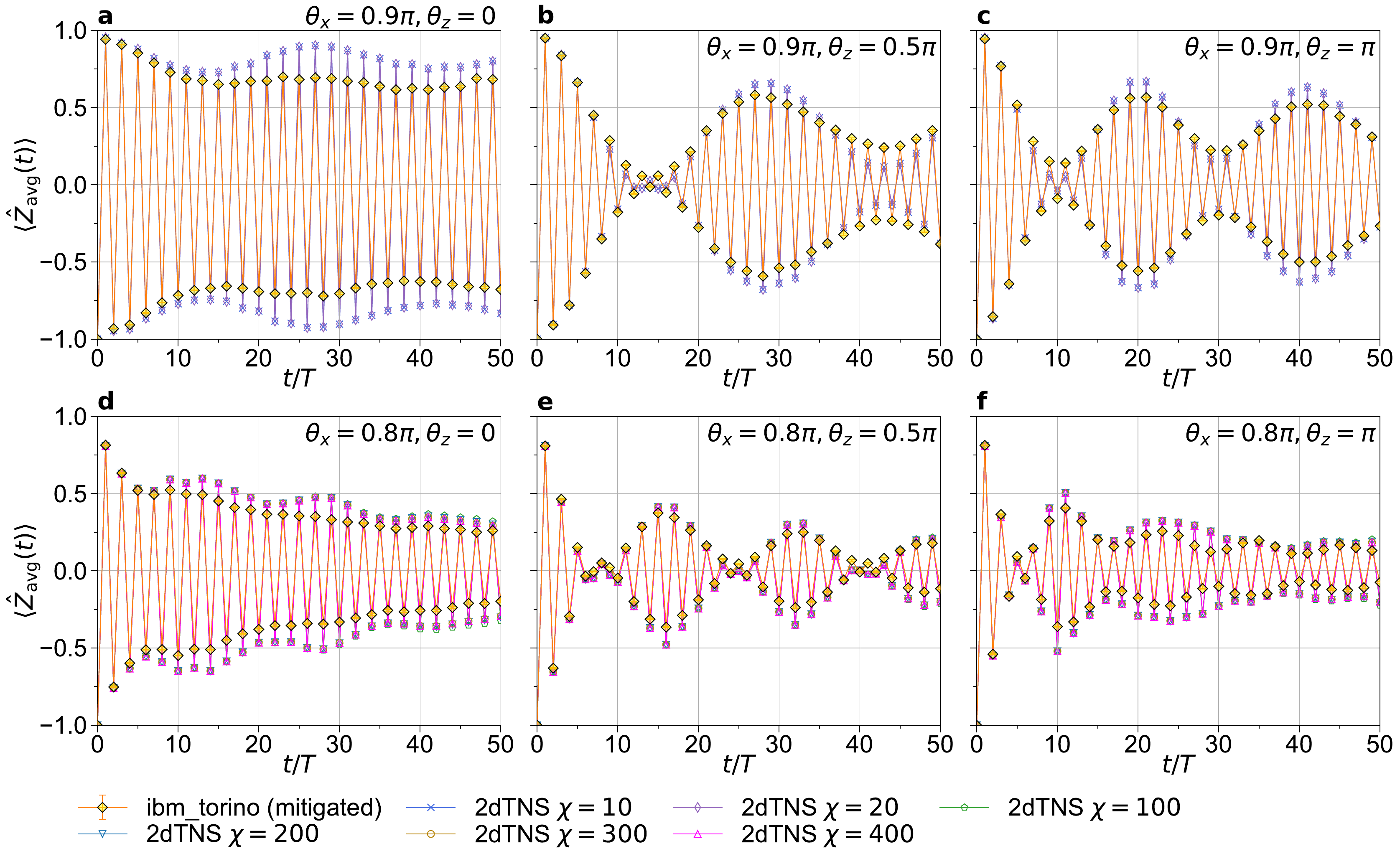}
\caption{
Enlarged plots of Fig.~\ref{fig:L133_peps}, focusing on the time evolution for up to 50 time steps.   
}
\label{fig:L133_peps_en}
\end{figure*}

\begin{figure*}[t]
\includegraphics[width=1.0\textwidth]{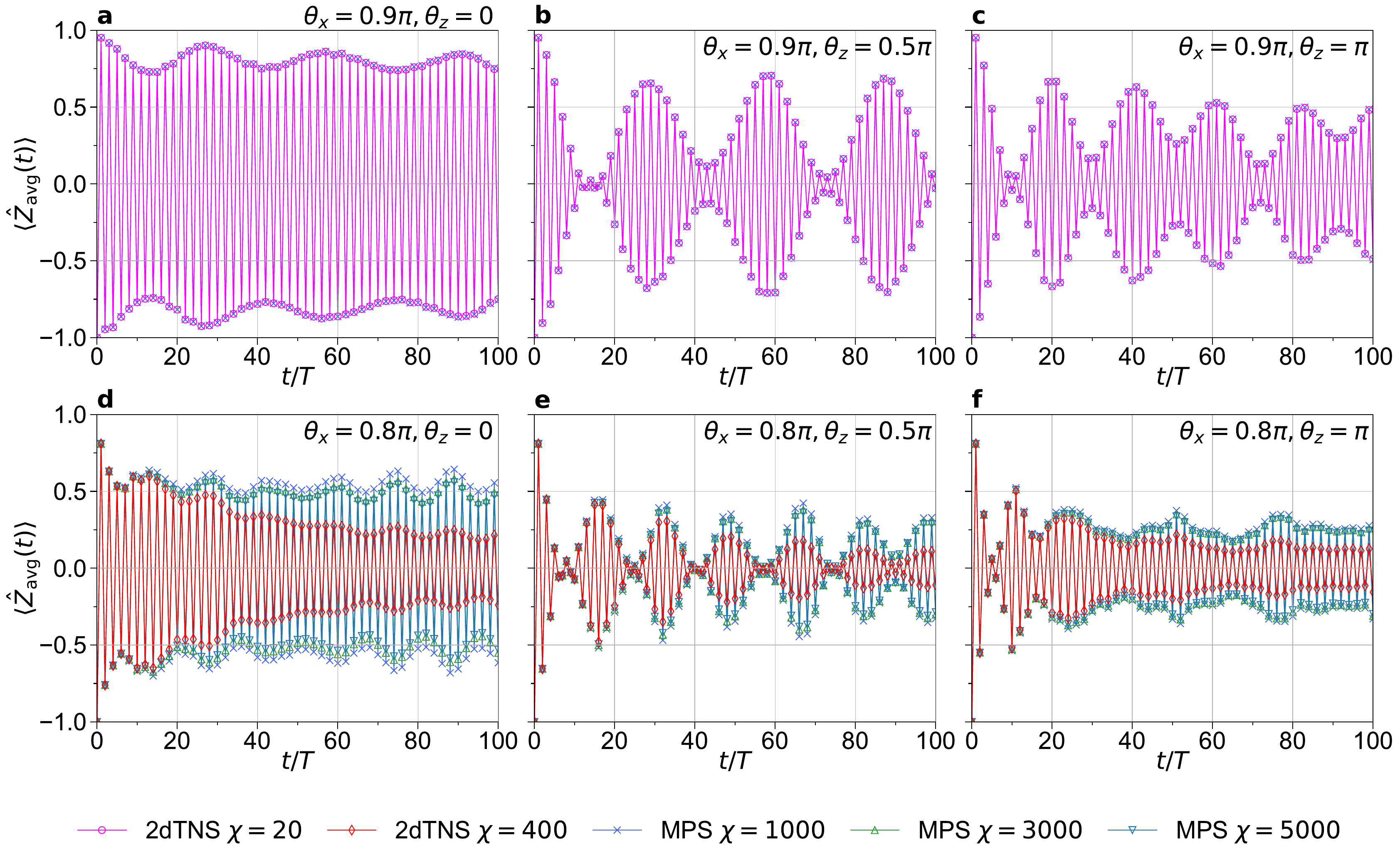}
\caption{
Comparison of the results for $\langle \avg{Z}(t) \rangle$ on the heavy-hexagonal lattice of $L=133$ qubits obtained by the 2dTNS method with a given bond dimension ($\chi=20$ in panels \cap{a}-\cap{c} 
and 400 in panels \cap{d}-\cap{f}) 
and the MPS method with various bond dimensions ($\chi=1000$, 3000, and 5000). 
The parameters $(\theta_x,\theta_z)$ are
\cap{a} $(0.9\pi,0)$, 
\cap{b} $(0.9\pi,0.5\pi)$, 
\cap{c} $(0.9\pi,\pi)$, 
\cap{d} $(0.8\pi,0)$, 
\cap{e} $(0.8\pi,0.5\pi)$, and 
\cap{f} $(0.8\pi,\pi)$.
}
\label{fig:L133_mps}
\end{figure*}

\begin{figure*}[t]
\includegraphics[width=1.0\textwidth]{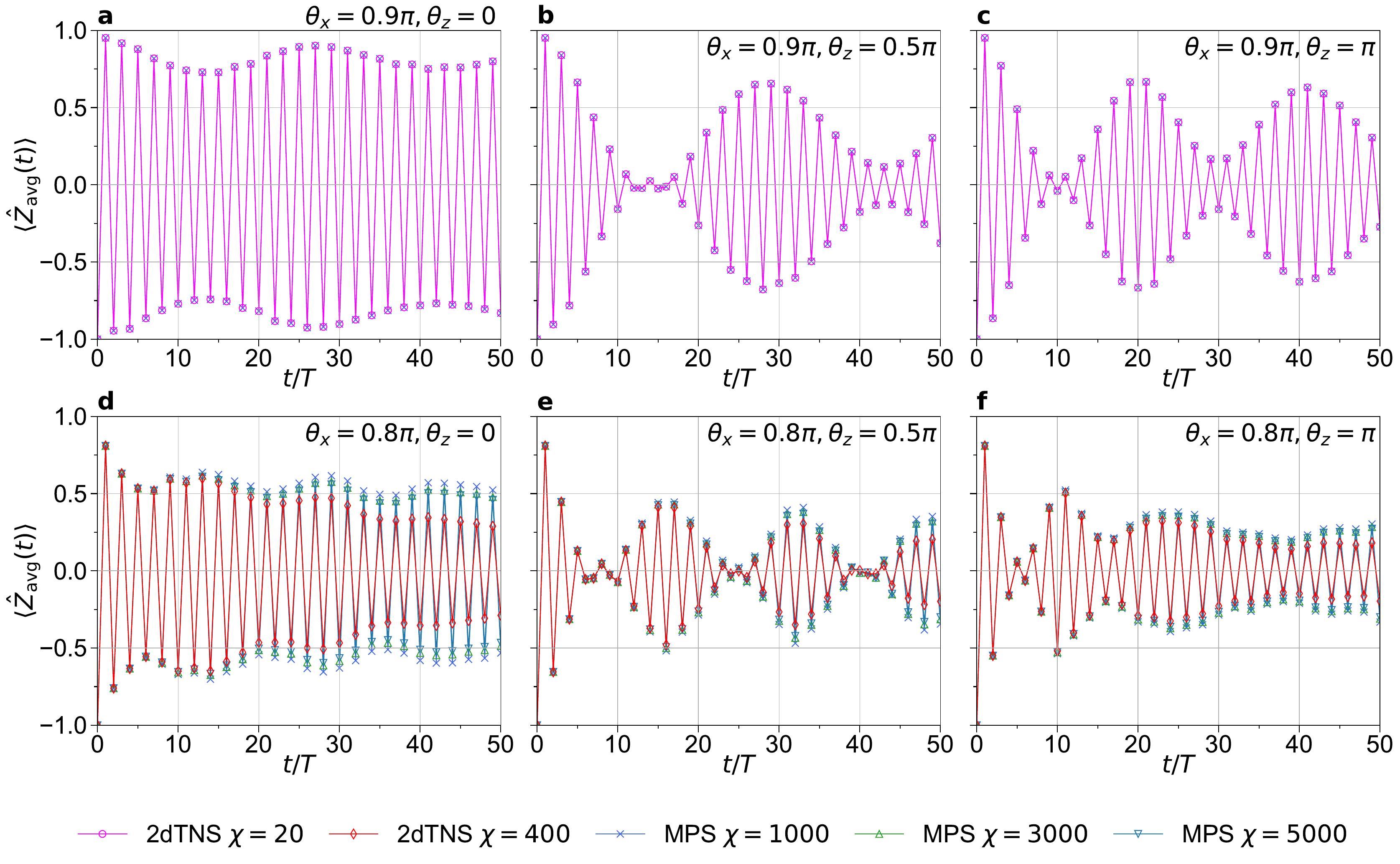}
\caption{
Enlarged plots of Fig.~\ref{fig:L133_mps}, focusing on the time evolution 
for up to 50 time steps. 
}
\label{fig:L133_mps_en}
\end{figure*}

In Fig.~2d of the main text, we demonstrated that the averaged magnetisation $\langle \avg{Z}(t) \rangle$ obtained from {\torino} with the error mitigation agrees well with those of classical 2dTNS simulations at $(\theta_{x},\theta_{z})=(0.8\pi,0.5\pi)$ for the $L=133$ system over 50 time steps. 
In Figs.~\ref{fig:L133_peps} and \ref{fig:L133_peps_en}, we extend this comparison to other parameters over 100 time steps. 
As shown in Figs.~\ref{fig:L133_peps}a-\ref{fig:L133_peps}c and \ref{fig:L133_peps_en}a-\ref{fig:L133_peps_en}c, when $\theta_{x}=0.9\pi$, the averaged magnetisation $\langle \avg{Z}(t) \rangle$ obtained by the 2dTNS method with a relatively small bond dimension $\chi=20$ already converges sufficiently. 
This implies that the 2dTNS results for these parameter sets are reliable even for up to 100 time steps. 
Indeed, this parameter region exhibits limited entanglement growth over time steps, as indicated by the agreement between the MPS results with relatively small bond dimensions and the 2dTNS results (see Figs.~\ref{fig:L133_mps}a-\ref{fig:L133_mps}c and \ref{fig:L133_mps_en}a-\ref{fig:L133_mps_en}c).
Moreover, it is particularly remarkable that despite the simple error mitigation protocol introduced in Eq.~(4) of the main text, the error-mitigated $\langle \avg{Z}(t) \rangle$ obtained from {\torino} are in relatively 
good alignment with the 2dTNS results (see Figs.~\ref{fig:L133_peps}a-\ref{fig:L133_peps}c and \ref{fig:L133_peps_en}a-\ref{fig:L133_peps_en}c).

On the other hand, as shown in Figs.~\ref{fig:L133_peps}d-\ref{fig:L133_peps}f and \ref{fig:L133_peps_en}d-\ref{fig:L133_peps_en}f, the convergence of the 2dTNS results for 
$\theta_{x}=0.8\pi$ is slow with increasing the bond dimension, especially for long time steps, as anticipated from the results for the $L=28$ system 
shown in Figs.~\ref{fig:L28_peps}d-\ref{fig:L28_peps}df. 
This parameter region is close to the crossover boundary between the DTC and IM-DTC responses and the thermalised regime, where extensive entanglement growth is expected. 
Due to the substantial increase in entanglement, the performance of tensor-network simulations is significantly degraded, resulting in a noticeable discrepancy between MPS and 2dTNS as shown in Figs.~\ref{fig:L133_mps}d-\ref{fig:L133_mps}f and \ref{fig:L133_mps_en}d-\ref{fig:L133_mps_en}f.
Although the classical 2dTNS simulations may not have yet converged in these parameter sets for $t\agt50$, we still observe that the error-mitigated $\langle \avg{Z}(t) \rangle$ obtained from {\torino} align well with the 2dTNS results 
(see Figs.~\ref{fig:L133_peps}d-\ref{fig:L133_peps}f and \ref{fig:L133_peps_en}d-\ref{fig:L133_peps_en}f). 
Given that this is the parameter region where classical simulations face challenges, it underscores the value of quantum computers precisely in this parameter region.

\begin{figure*}[t]
\includegraphics[width=1.0\textwidth]{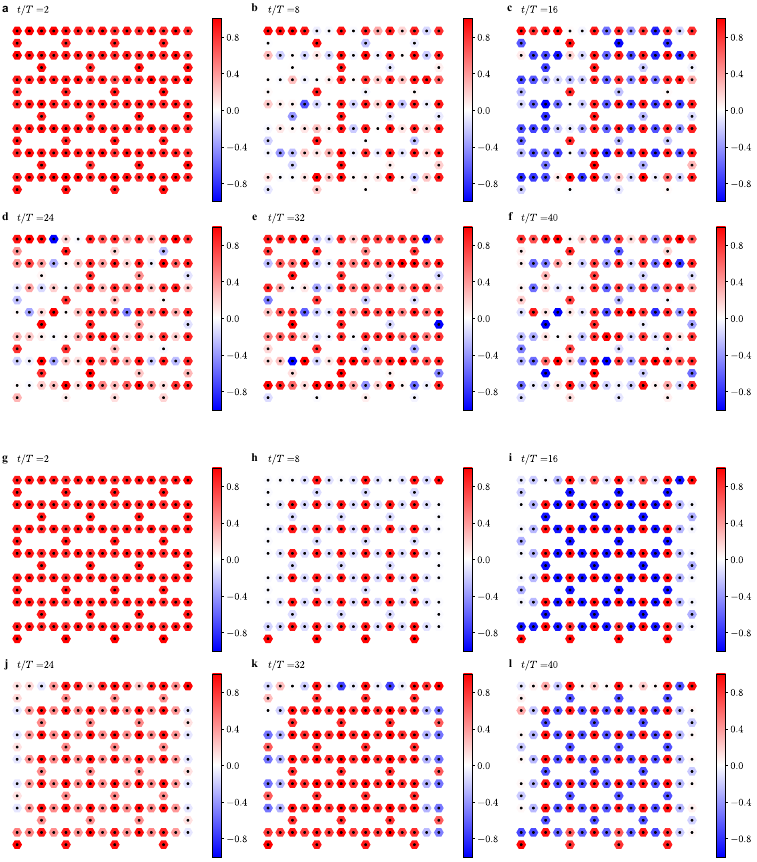}
\caption{Time evolution of local expectation values $\langle Z_{j}(t) \rangle$ at $t/T=2$, 8, 16, 24, 32, and 40 on the heavy-hexagonal lattice of $L=133$ qubits with the parameter $(\theta_{x},\theta_{z})=(0.9\pi,0.5\pi)$. 
		\cap{a}-\cap{f} Error-mitigated expectation values obtained using {\torino}.
		\cap{g}-\cap{l} The corresponding results obtained by the MPS method with the bond dimension $\chi=1000$.
		Here, the initial state is prepared as a product state with all qubits set to $|0\rangle$, representing a fully-polarised state.
  }
\label{fig:L133_mps_snap}
\end{figure*}

\begin{figure*}[t]
\includegraphics[width=1.0\textwidth]{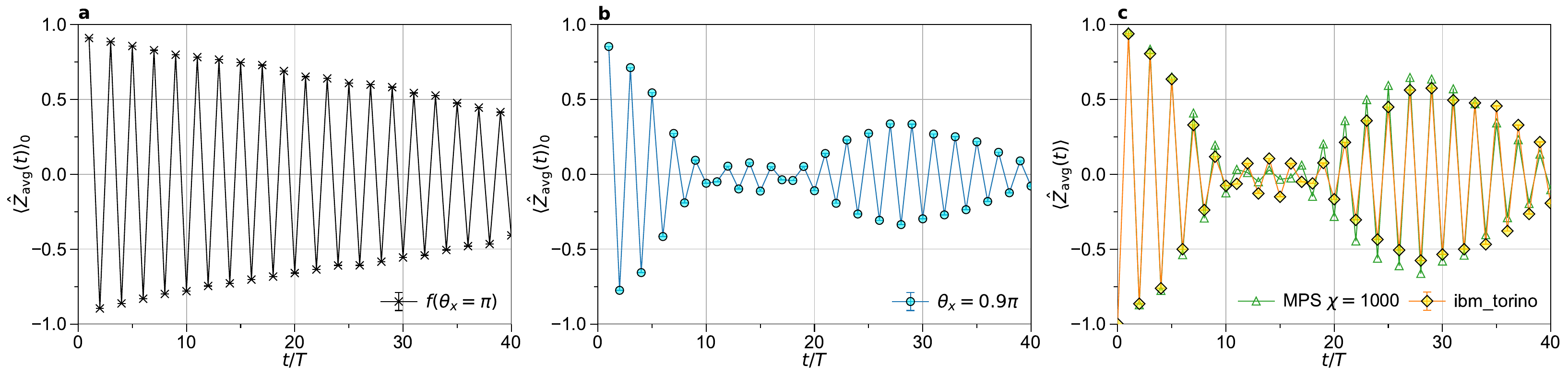}
\caption{
    \cap{a} Raw data of the averaged magnetisation $\langle \avg{Z}(t)\rangle_{0,\theta_x=\pi}$ for the heavy-hexagonal lattice of 
    $L=133$ qubits with the parameter 
    $(\theta_x,\theta_z)=(\pi,0.5\pi)$ obtained using {\torino}.
    \cap{b} Same as \cap{a}, but with the parameter $(\theta_x,\theta_z)=(0.9\pi,0.5\pi)$.
    \cap{c} Error-mitigated data (yellow diamonds) of the averaged magnetisation $\langle \avg{Z}(t)\rangle$ for the heavy-hexagonal lattice of 
    $L=133$ qubits with the parameter
    $(\theta_x,\theta_z)=(0.9\pi,0.5\pi)$. 
    The results obtained by the MPS method with the bond dimension $\chi=1000$ are also shown with green triangles. 
    Here, the initial state is prepared as a product state with all qubits set to $|0\rangle$, i.e., a fully-polarised state. 
     }
\label{fig:comp_fm}
\end{figure*}

In Figs.~\ref{fig:L133_mps_snap}a-\ref{fig:L133_mps_snap}f, we show the spatial distribution of the error-mitigated magnetisation $\langle \hat{Z}_{j}(t) \rangle$ at $t/T=2$, 16, 24, 32, and 40, obtained using {\torino} for the $L=133$ system with the parameter $(\theta_{x},\theta_{z})=(0.9\pi,0.5\pi)$. 
Here, the same error mitigation protocol as defined in Eq.~(4) of the main text is employed, except that the denominator on the right hand side of Eq.~(4) is replaced by the spatial average of the magnetisation over the entire system.
These results are compared with those obtained using the MPS method in Figs.~\ref{fig:L133_mps_snap}g-\ref{fig:L133_mps_snap}l. 
The initial state $|\psi(0)\rangle$ is prepared as a product state with all qubits set to $|0\rangle$, representing a fully polarised state. 
As shown in Fig.~\ref{fig:comp_fm}c, we also observe a longer-period 
oscillation in the averaged magnetisation $\langle \avg{Z}(t) \rangle$ 
upon introducing non-zero $\theta_{z}$, indicating an IM-DTC response, despite the initial state differing from that used elsewhere in this study.
Interestingly, the period of this oscillation closely resembles that observed in Figs.~\ref{fig:L133_peps}b and \ref{fig:L133_peps_en}b, where the initial state is set differently. 
Furthermore, we find in Fig.~\ref{fig:comp_fm}c that the error-mitigated data are in good agreement with the results of MPS over 40-time steps.

Now, it is interesting to investigate the nature of the time-evolved state at characteristic times. 
As shown in Fig.~\ref{fig:comp_fm}c, the nodes of the longer-period oscillation occur at $t/T\simeq14$ and slightly beyond $40$, which are similar to those observed in Fig.~\ref{fig:L133_peps_en}b.
At times near these nodes, specifically $t/T=16$ and $40$ as shown in Figs.~\ref{fig:L133_mps_snap}i and \ref{fig:L133_mps_snap}l, respectively, we observe that the time-evolved state $|\psi(t)\rangle$ exhibits a Neel-like structure in the MPS simulations. 
Even in {\torino}, we observe, at least partially, a similar Neel-like structure, as shown in Figs.~\ref{fig:L133_mps_snap}c and \ref{fig:L133_mps_snap}f. 
Conversely, at times near the antinodes, such as $t/T=2$, $24$, and $32$ shown in Figs.~\ref{fig:L133_mps_snap}g, \ref{fig:L133_mps_snap}j, 
and \ref{fig:L133_mps_snap}k, respectively, the time-evolved state $|\psi(t)\rangle$ exhibits a ferromagnetic (FM)-like structure in the MPS simulations.
Similarly, in {\torino}, we observe a comparable FM-like structure, albeit partially, as shown in Figs.~\ref{fig:L133_mps_snap}a, \ref{fig:L133_mps_snap}d, and \ref{fig:L133_mps_snap}e.

\section{Results for a one-dimensional lattice of $L=112$ qubits} \label{sec:1d}

While the emergence of a DTC in a one-dimensional system with short-range interactions necessitates many-body localisation, exploring the one-dimensional case driven by the same Floquet operator $\hat{U}_{\rm F}$ in Eq.~(2) remains valuable. 
This investigation serves to highlight differences from the two-dimensional case examined throughout this paper and allows for further assessment of the reliability of results obtained using {\torino}, along with the error-mitigation protocol introduced in Eq.~(4).

\begin{figure}[htbp]
\includegraphics[width=0.45\textwidth]{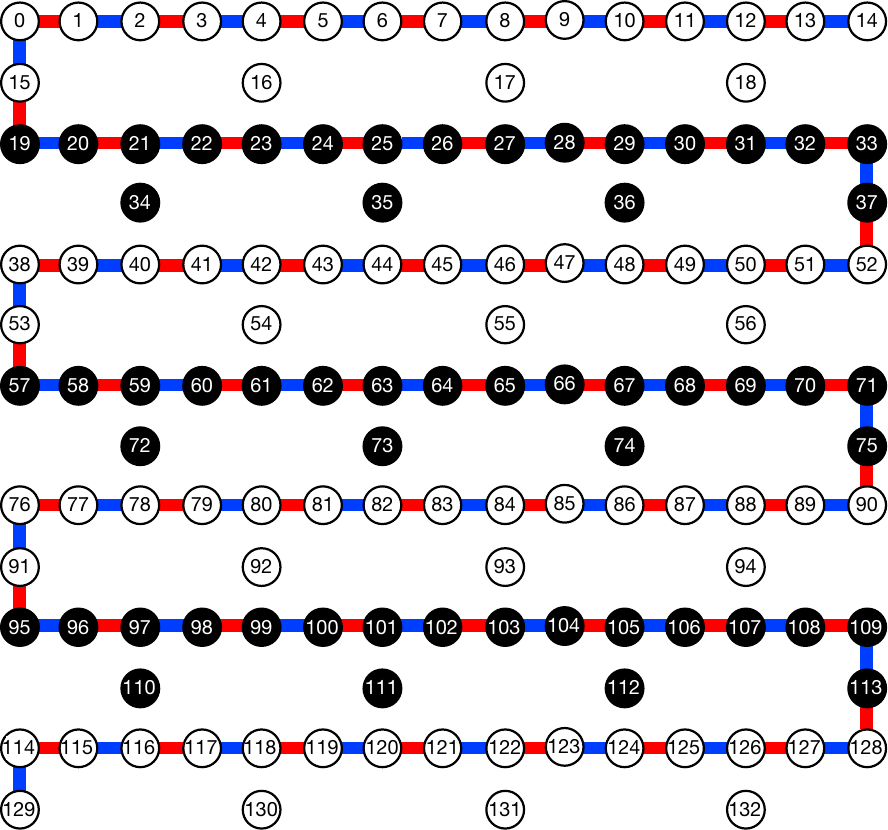}
\caption{
   A one-dimensional lattice of $L=112$ qubits embedded in the heavy-hexagonal lattice geometry of the IBM Quantum Heron processor, {\torino}. 
   The qubits forming the one-dimensional lattice are connected by red and blue edges. 
   When the quantum device is utilised, $\hat{R}_{Z_{i}Z_{j}} (\theta_{J})$ gates in the single-cycle Floquet operator $\hat{U}_{\rm F}$ applied on red or blue bonds are operated in parallel. 
  The initial state is prepared as a product state with qubits indicated by white (black) circles set to $|0\rangle$ ($|1\rangle$). 
  }
\label{fig:geometry_1d}
\end{figure}

A one-dimensional lattice of $L=112$ qubits can be incorporated into the heavy-hexagonal lattice geometry of the IBM Heron processor, {\torino}, as depicted in Fig.~\ref{fig:geometry_1d}. In this configuration, qubits constituting the one-dimensional lattice are interconnected by red and blue edges. 
The two-qubit $\hat{R}_{Z_{i}Z_{j}} (\theta_{J})$ gates in the single-cycle Floquet operator $\hat{U}_{\rm F}$ are applied on red or blue bonds concurrently in the quantum circuit. 
The initial state $|\psi(0)\rangle$ for the time evolution is prepared as a product state, forming a domain-wall configuration of $|0\rangle$'s and $|1\rangle$'s, represented by white and black circles, respectively, in Fig.~\ref{fig:geometry_1d}. 
We evaluate the time evolution of the averaged magnetisation ${\avg{Z}}(t)$, defined in Eq.~(3), over the same 
set $A$ of qubits as in the heavy-hexagonal lattice of $L=133$ qubits, 
i.e., $A=\{ 57, 58, 59, 60, 61, 62, 63, 64, 65, 66, 67, 68, 69, 70, 71 \}$ 
(the qubit labels are indicated in Fig.~\ref{fig:geometry_1d}).

\begin{figure}[t]
\includegraphics[width=1\textwidth]{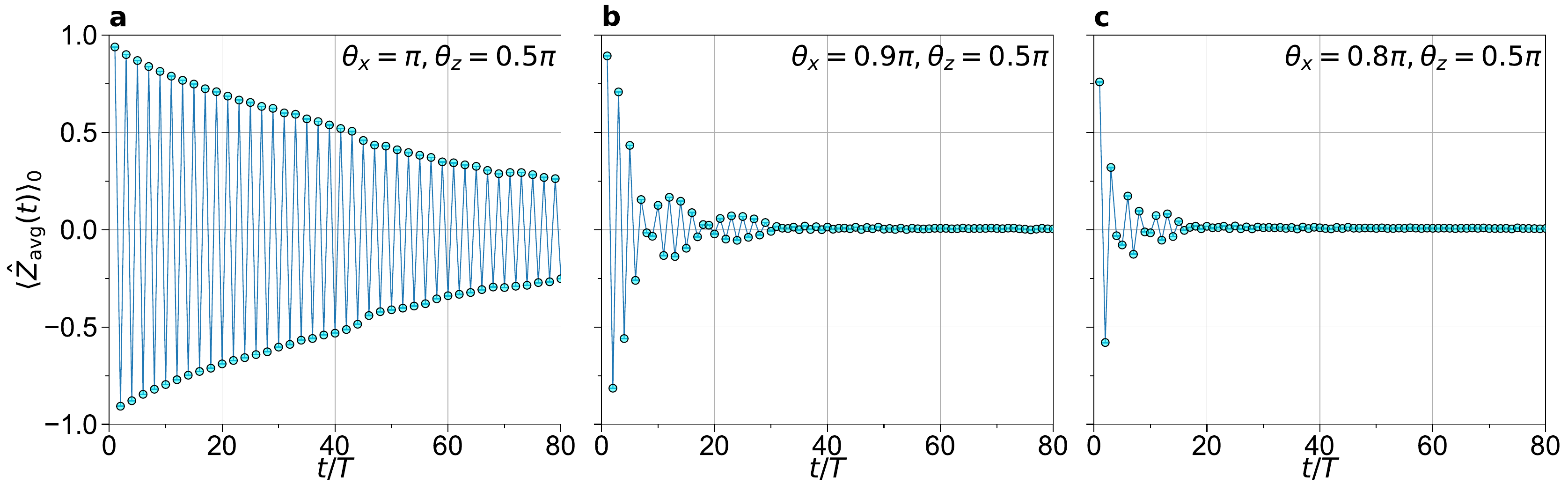}
\caption{
Error-unmitigated raw data $\langle \avg{Z}(t) \rangle_{0}$ for the one-dimensional lattice of $L=112$ qubits obtained using {\torino} (see Fig.~\ref{fig:geometry_1d}). 
The parameters $(\theta_x,\theta_z)$ are 
\cap{a} $(\pi,0.5\pi)$, 
\cap{b} $(0.9\pi,0.5\pi)$, and 
\cap{c} $(0.8\pi,0.5\pi)$. 
}
\label{fig:L112_raw}
\end{figure}

Figures~\ref{fig:L112_raw}b and \ref{fig:L112_raw}c show the error-unmitigated raw data $\langle \avg{Z}(t) \rangle_0$ obtained using {\torino} for the one-dimensional system of $L=112$ qubits with the parameters 
$(\theta_x,\theta_z)=(0.9\pi,0.5\pi)$ and $(0.8\pi,0.5\pi)$, respectively. 
Additionally, Fig.~\ref{fig:L112_raw}a shows the error-unmitigated raw data $\langle \avg{Z}(t) \rangle_0$ for the same system but with $(\theta_x,\theta_z)=(\pi,0.5\pi)$, which are utilised to mitigate errors for $\langle \avg{Z}(t) \rangle_0$ with $(\theta_x,\theta_z)=(0.9\pi,0.5\pi)$ and $(0.8\pi,0.5\pi)$ shown in Figs.~\ref{fig:L112_raw}b and \ref{fig:L112_raw}c, respectively. 
From the value of $|\langle \avg{Z}(t=80T) \rangle_0|$ shown in Fig.~\ref{fig:L112_raw}a, we can estimate the effective quantum circuit volume $v_{\rm eff}\simeq 350$ at $t/T=80$, which is significantly smaller than the quantum circuit volume $v=8880$ (the number of CZ gates in the quantum circuit increases by 111 for each operation of $\hat{U}_{\rm F}$ in the one-dimensional system).

\begin{figure}[t]
\includegraphics[width=1\textwidth]{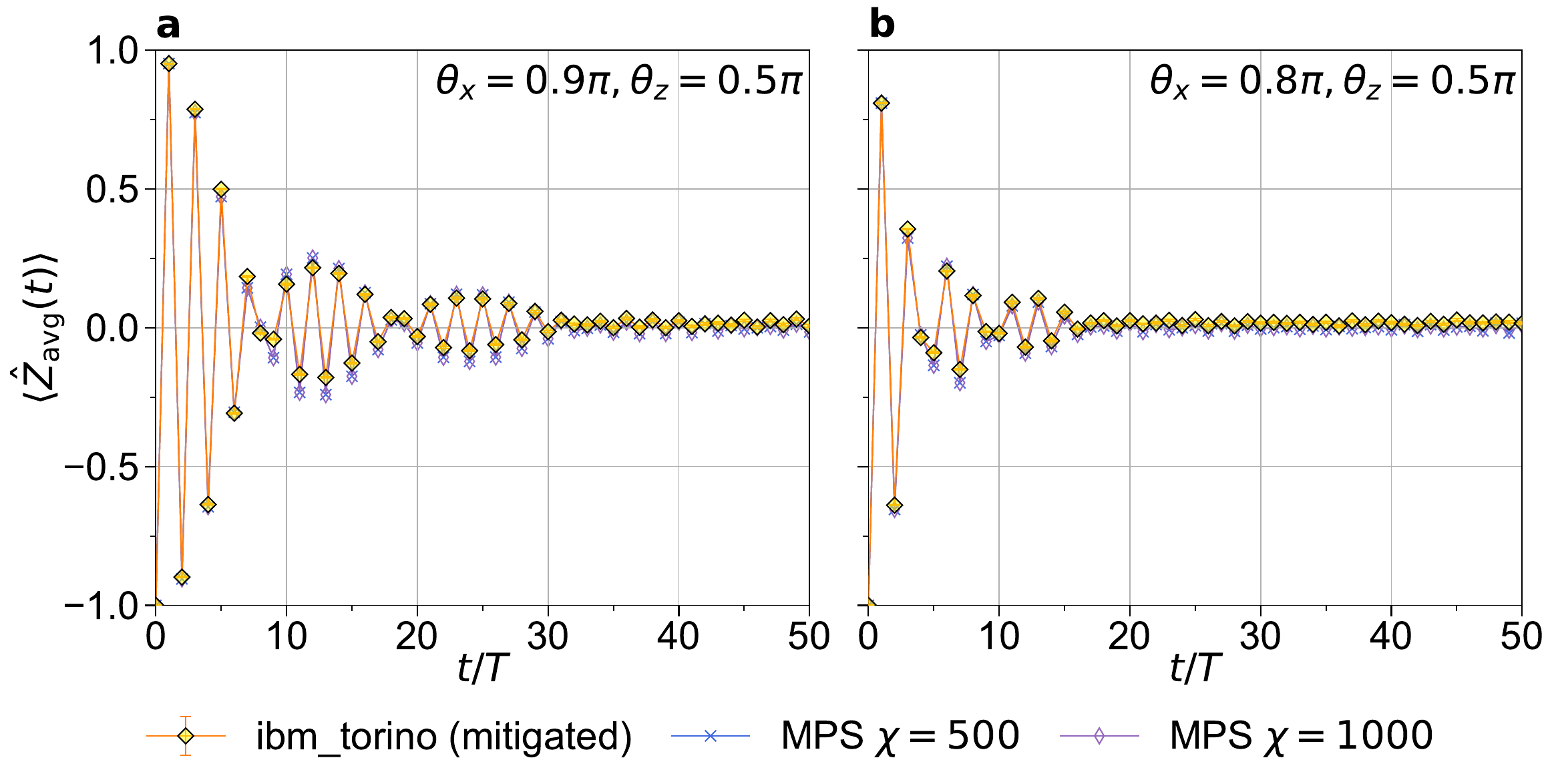}
\caption{
Error-mitigated data $\langle \avg{Z}(t) \rangle$ (yellow diamonds) are compared with the results obtained by the MPS simulations with the bond dimensions $\chi=500$ and 1000 (shown as blue crosses and purple diamonds, respectively) for the one-dimensional lattice of $L=112$ qubits.
The parameters $(\theta_x,\theta_z)$ are \cap{a} $(0.9\pi,0.5\pi)$ and \cap{b} $(0.8\pi,0.5\pi)$. 
}
\label{fig:L112_mit}
\end{figure}

The error-mitigated data $\langle \avg{Z}(t) \rangle$ for the one-dimensional system of $L=112$ qubits with the parameters $(\theta_x,\theta_z)=(0.9\pi,0.5\pi)$ and $(0.8\pi,0.5\pi)$ are shown in Fig.~\ref{fig:L112_mit}. 
In sharp contrast to the case for the heavy-hexagonal lattice of $L=133$ qubits, the period-doubling oscillations observed in the one-dimensional system are much weaker against the deviation of $\theta_{x}$ from $\pi$. 
We find that the signals with these oscillations decay rather quickly in time steps for other parameter sets of $(\theta_x,\theta_z)$ in one dimension, even when $\theta_z=0.9$. 
These observations are consistent with the expectation that, in one dimension with short-range interactions and without disorder, neither a long-lived DTC nor an IM-DTC response is stabilised.

Moreover, in Fig.~\ref{fig:L112_mit}, the results for 
$\langle \avg{Z}(t) \rangle$ obtained using {\torino} are compared with those calculated by the MPS method with the bond dimensions $\chi=500$ and 1000, exhibiting the converged values.
Remarkably, despite the simplicity of the error-mitigation protocol introduced, these two results obtained using {\torino} and the MPS method show excellent agreement, confirming the reliability of the results obtained using {\torino}. 


\section{The role of Ising interaction}
\begin{figure*}[t]
\includegraphics[width=1.0\textwidth]{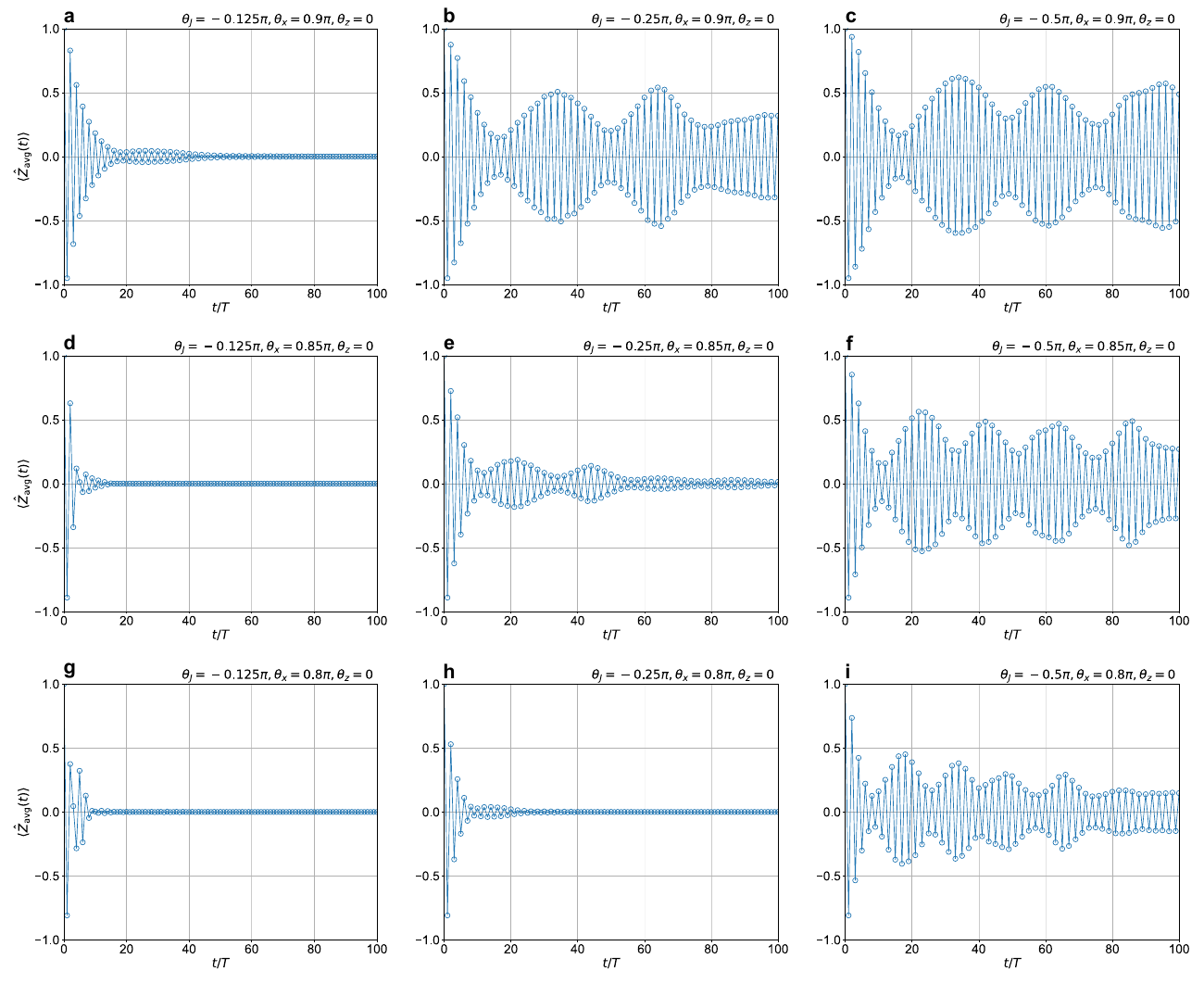}
\caption{
Average magnetization $\langle \avg{Z}(t) \rangle$ for the heavy-hexagonal lattice of $L=28$ qubits obtained using the state-vector method.
The parameter $\theta_x$ is set to $0.9\pi$ for \cap{a}--\cap{c}, $0.85\pi$ for \cap{d}--\cap{f}, and $0.8\pi$ for \cap{g}--\cap{i}, while the parameter $\theta_z$ is fixed at 0. 
The parameter $\theta_J$ of the Ising interaction is $-0.125\pi$ in \cap{a}, \cap{d}, and \cap{g}, $-0.25\pi$ in \cap{b}, \cap{e}, and \cap{h}, and $-0.5\pi$ in \cap{c}, \cap{f}, and \cap{i}. 
Here, the initial state $|\psi(0)\rangle$ is prepared as a product state with all qubits set to $|0\rangle$, i.e., a fully-polarised state, and the magnetisation $\langle \avg{Z}(t) \rangle$ is averaged over all qubits.}
\label{fig:interaction}
\end{figure*}

In this section, we demonstrate that the Ising interaction $J$ plays an important role in stabilising DTCs.
Figure~\ref{fig:interaction} shows the averaged magnetisation $\langle \avg{Z}(t) \rangle$ with different values of $\theta_J$ and $\theta_x$, setting $\theta_z=0$, for the heavy-hexagonal lattice of $L=28$ qubits (see Fig.~1a), obtained using the state-vector method. 
Here, we initialise the state $|\psi(0)\rangle$ as a product state with all qubits set to $|0\rangle$ and the magnetisation $\langle \avg{Z}(t) \rangle$ is averaged over all qubits.
We find that the period-doubling oscillations are quite stable at $\theta_J=-0.5\pi$, which is the parameter used in all other cases of this paper and where the DTC and IM-DTC signals are also clearly observed experimentally in ibm\_torino.
In contrast, for small $\theta_J$, magnetisation oscillations are fragile against the deviation of $\theta_x$ from $\pi$, leading to a rapid decay of oscillations.
This behavior indicates that the DTCs observed on the heavy-hexagonal lattice are supported by strong interactions and are realised within a finite-frequency regime.

\section{Error mitigation based on a global depolarizing noise model}

In this section, we provide a detailed description of our error mitigation protocol.
When a qubit is depolarised into a fully mixed state $I/2$ with probability $p$, the depolarising noise channel modifies the qubit's density matrix $\rho$ as~\cite{S:Nielsen_Chuang_2010}
\begin{align}\label{eq:dp}
    \rho \rightarrow \mathcal{E}(\rho)=&(1-p)\rho + p(I/2).
\end{align}
Accordingly, in an $L$-qubit system, the expectation value of the $i$-th qubit, $\langle \hat{Z}_i(t) \rangle_\text{dp}$, under depolarizing noise is given by
\begin{align}\label{eq:dp1}
    \langle \hat{Z}_i(t) \rangle_\text{dp} \simeq e^{-tp_i} \langle \hat{Z}_i(t) \rangle,
\end{align}
where $p_i$ is the depolarising probability for the $i$-th qubit and $\langle \hat{Z}_i(t) \rangle$ represents the magnetisation without quantum noise.
The averaged expectation value defined in Eq.~(3) of the main text reads
\begin{align}
    \langle \avg{Z}(t) \rangle_\text{dp} =& \frac{1}{|A|}\sum_{j\in A} e^{-tp_j} \langle \hat{Z}_j(t) \rangle\\
    \simeq & \int dp f(p)e^{-tp}\langle \avg{Z}(t) \rangle,
\end{align}
where, in the second line, we have introduced a probability density function $f(p)$ to approximate the discrete sum. 
This approximation assumes a sufficiently large system size and averaging over a large number of qubits.
Under these conditions, it is expected that $\langle \hat{Z}_j \rangle$ does not depend strongly on each qubit. 
Assuming a Gaussian distribution $f(p)=\frac{1}{\sqrt{2\pi}\sigma}\exp\left[ -\frac{(p-p_0)^2}{2\sigma^2} \right]$ with mean $p_0$ and variance $\sigma^2$, we obtain $\langle \avg{Z}(t) \rangle_\text{dp} \simeq e^{-tp_0}\langle \avg{Z}(t) \rangle$.
Here, in this study, we adopt a global depolarising noise model, where the depolarisation probability $p$ is nearly uniform across qubits, implying $\sigma \sim 0$.
In the following, we argue that this assumption is reasonable under the unitary circuit of the kicked Ising model.

\begin{figure*}[t]
\includegraphics[width=1.0\textwidth]{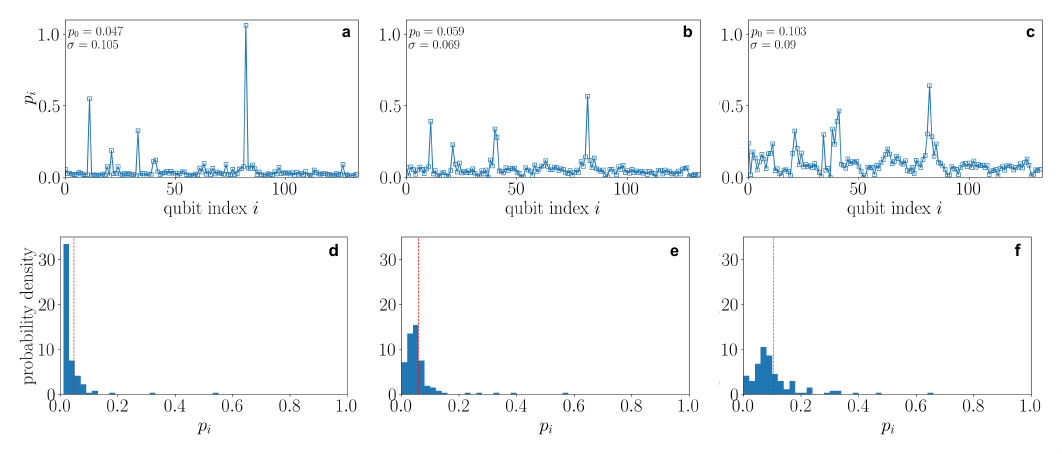}
\caption{Depolarisation probability $p_i$ for each qubit on ibm\_torino estimated for the kicked Ising model on the heavy-hexagonal lattice of $L=133$ qubits. 
        The model parameter is set to \cap{a} $\theta_x=\pi$, \cap{b} $0.9\pi$, and \cap{c} $0.8\pi$, with $\theta_z$ fixed at zero. 
        The corresponding values of $p_0$ and $\sigma$ are also shown for each $\theta_x$.
        Histograms of $p_i$ are shown for \cap{d} $\theta_x=\pi$, \cap{e} $0.9\pi$, and \cap{f} $0.8\pi$, where the red dashed lines indicate $p_0$.
        }
\label{fig:dp1}
\end{figure*}

In Fig.~\ref{fig:dp1}, we present the depolarisation probability $p_i$ for each qubit on ibm\_torino, estimated for the kicked Ising model on the heavy-hexagonal lattice of $L=133$ qubits.
To estimate $p_i$, we fit the function $e^{-tp_i} \langle \hat{Z}_i(t) \rangle$ to the experimental data $\langle \hat{Z}_i(t) \rangle_\text{dp}$, based on Eq.~(\ref{eq:dp1}).
Here, we use the 2dTNS simulation results for the noise-free magnetisation $\langle \hat{Z}_i(t) \rangle$ on the right-hand side of Eq.~(\ref{eq:dp1}), and the raw experimental data obtained from ibm\_torino for $\langle \hat{Z}_i(t) \rangle_\text{dp}$ on the left-hand side. 
The experiment was executed on July 29, 2024. 
Except for several noisy qubits, the values of $p_i$ exhibit only weak dependence on the qubit index.
This behavior is also reflected in the histograms showing the distribution of $p_i$ values. 
As seen in Figs.~\ref{fig:dp1}d--\ref{fig:dp1}f, it is therefore reasonable to approximate the depolarising probabilities by a qubit-independent value $p_0$.

\begin{figure*}[t]
\includegraphics[width=1.0\textwidth]{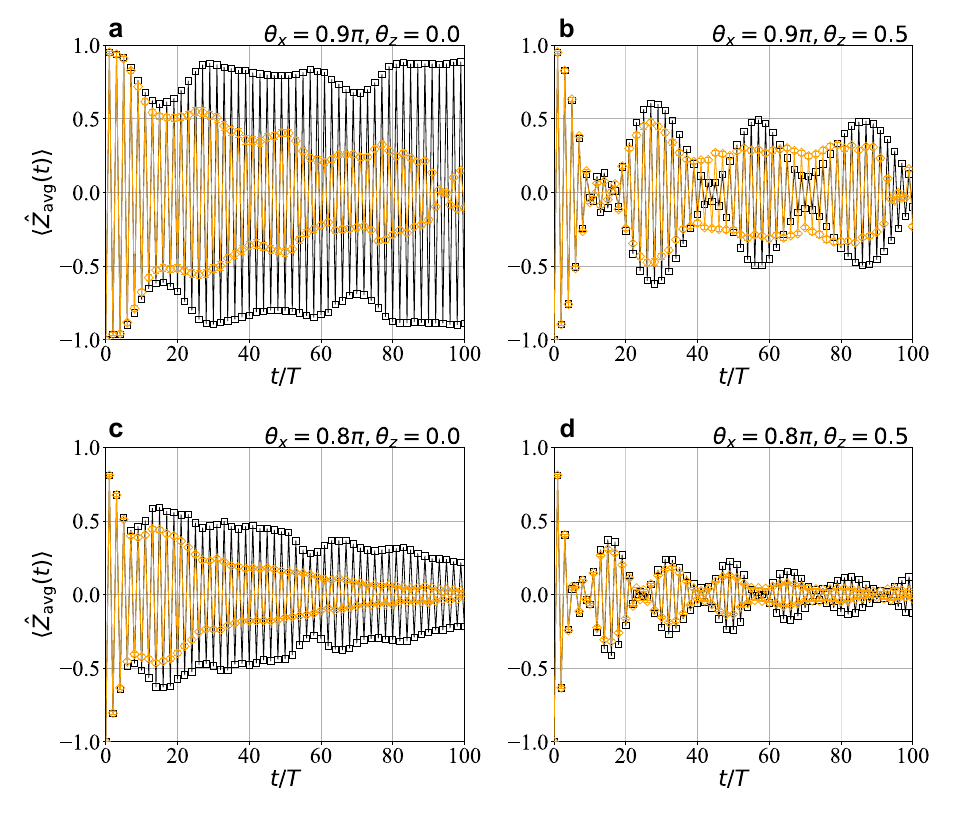}
\caption{
Error-mitigated average magnetisation $\langle \avg{Z}(t) \rangle$ 
(yellow diamonds) are compared with the noise-free results (black squares),  
both obtained by the classical state-vector method,  
for the heavy-hexagonal lattice of $L=28$ qubits. 
Quantum noise is introduced via the depolarising channels defined in 
Eqs.~(\ref{eq:depolarizingnoise1}) and (\ref{eq:depolarizingnoise2}), with averages taken over 100 independent noise realizations using 
$p_2=10p_1=0.01$. 
The parameters $(\theta_x,\theta_z)$ are 
\cap{a} $(0.9\pi,0)$, 
\cap{b} $(0.9\pi,0.5\pi)$, 
\cap{c} $(0.8\pi,0)$, and
\cap{d} $(0.8\pi,0.5\pi)$.
The error bars indicated represent the propagated error in the quantities 
in the numerator and the denominator of Eq.~(4).}
\label{fig:dp100}
\end{figure*}

\begin{figure*}[t]
\includegraphics[width=1.0\textwidth]{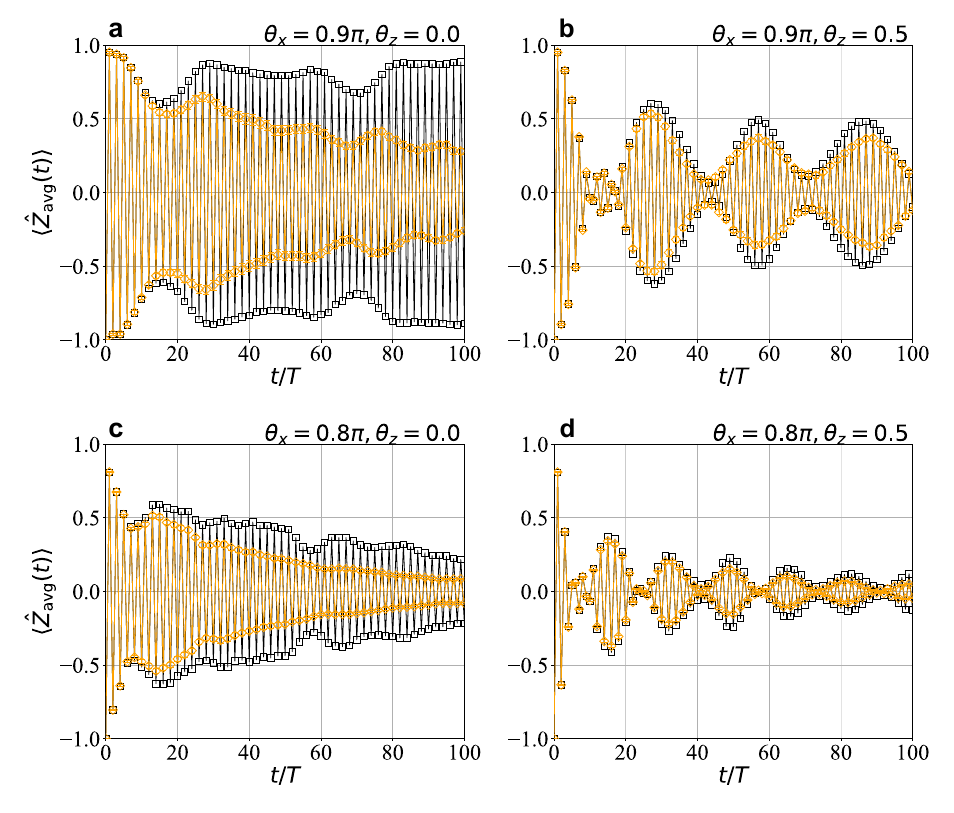}
\caption{
Same as Fig.~\ref{fig:dp100}, but with $p_2=10p_1=0.005$.}
\label{fig:dp200}
\end{figure*}

To further justify our error mitigation protocol based on the global depolarizing noise model, we perform state-vector simulations for the heavy-hexagonal lattice of $L=28$ qubits. In these simulations, quantum noise is incorporated by applying one- and two-qubit depolarising channels to the system's density matrix $\rho$ as~\cite{S:Nielsen_Chuang_2010,S:Ippoliti2021}
\begin{align}
    \mathcal{E}_i^{(1q)}(\rho)=&(1-p_1)\rho + \frac{p_1}{3}\sum_{\alpha \neq 0} \sigma_{\alpha,i} \rho \sigma_{\alpha,i}\label{eq:depolarizingnoise1}\\
    \mathcal{E}_{ij}^{(2q)}(\rho)=&(1-p_2)\rho + \frac{p_2}{15}\sum_{\alpha,\,\beta}\nolimits' \sigma_{\alpha,i}\sigma_{\beta,j} \rho \sigma_{\alpha,i}\sigma_{\beta,j}\label{eq:depolarizingnoise2},
\end{align}
where $\sigma_{\alpha,i}$ denotes the Pauli operators acting on qubit $i$, with $\sigma_{\alpha,i}=I$, $X_i$, $Y_i$, and $Z_i$ for $\alpha=0$, 1, 2, and 3, respectively, and the primed sum in Eq.~(\ref{eq:depolarizingnoise2}) exclude the identity pair $(\alpha,\beta)=(0,0)$.
Each single-qubit gate acting on qubit $i$ is followed by the channel $\mathcal{E}_i^{(1q)}$, and each two-qubit gate acting on qubits $i$ and $j$ is followed by $\mathcal{E}_{ij}^{(2q)}$.
We assume qubit-independent noise parameters and set $p_1=p_2/10$, which approximately matches the noise characteristics of the ibm\_torino device.
Under this noise model, we evaluate the time evolution of the averaged magnetisation $\langle \avg{Z}(t)\rangle$, as shown in Figs.~\ref{fig:dp100} and ~\ref{fig:dp200} for $p_2=0.01$ and $p_2=0.005$, respectively.
The same error mitigation protocol described in Eq.~(4) of the main text is applied but with quantum noises generated according to the depolarising channels defined in Eqs.~(\ref{eq:depolarizingnoise1}) and (\ref{eq:depolarizingnoise2}). We average over 100 independent noise realizations. 
More precisely, for each time step, we take sample averages over 100 noise realizations separately for the numerator and denominator in Eq.~(4), using the same set of noise realizations for both. The final error-mitigated value is then obtained by taking the ratio of these sample-averaged quantities.
Comparing Figs.~\ref{fig:dp100} and \ref{fig:dp200} with Fig.~\ref{fig:L28_mit}, we find that $p_2=0.01$ is too large to accurately characterise the quantum noises in ibm\_torino.
On the other hand, the behavior of $\langle \avg{Z}(t)\rangle$ obtained with $p_2=0.005$ (see Fig.~\ref{fig:dp200}) closely resembles that observed in the experimental results (see Fig.~\ref{fig:L28_mit}), at least for $t/T<50$.
Note that the average error rate of CZ gate on ibm\_torino is approximately 0.006.

From the above discussions, we conclude that the global depolarising noise model provides an effective description of quantum noise in the observation of $\langle \avg{Z}(t)\rangle$ for the kicked Ising model. 
The effectiveness of this model for mitigating quantum noise has also been pointed out in Ref.~\cite{S:Vovrosh2021}.
In quantum simulations of the kicked Ising model with large system size $L$, the number of gates per circuit layer becomes substantial, making the depolarising noise model a suitable effective approximation for characterising the accumulated noise.

\begin{figure*}[t]
\includegraphics[width=1.0\textwidth]{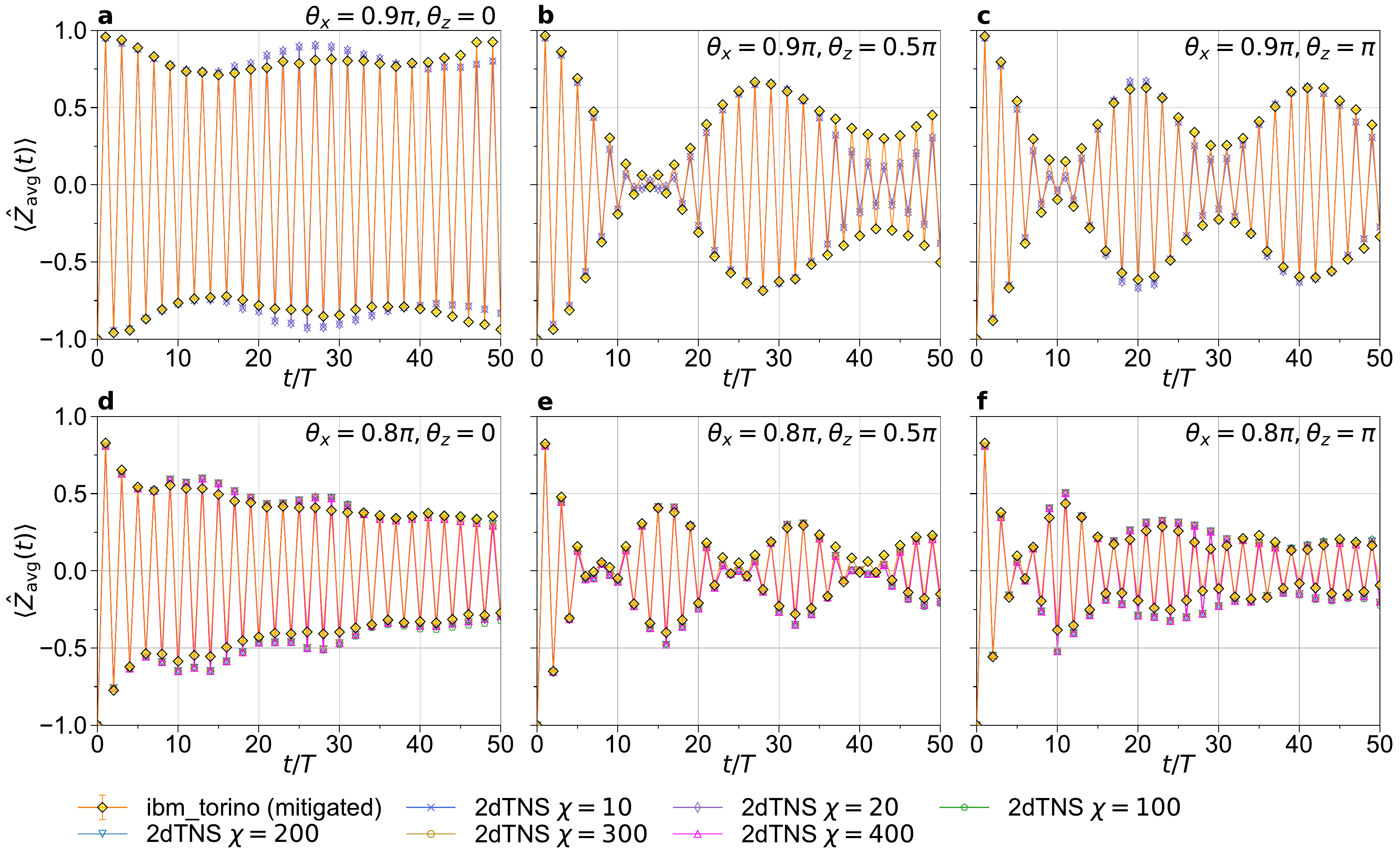}
\caption{
Same as Fig.~\ref{fig:L133_peps_en}, but using the modified error mitigation protocol defined in Eq.~(\ref{eq:norm-mod}).}
\label{fig:figs8_5over4}
\end{figure*}

The depolarising noise model is based on the simplifying assumption that all types of Pauli errors act equivalently. This is one reason why it is often justified in random unitary circuits and/or the deep regions of unitary circuits.
In the kicked Ising model, particularly for $\theta_x \sim \pi$, the effect of the depolarising noise defined in Eq.~(\ref{eq:depolarizingnoise2}) is expected to be anisotropic, assuming that the initial state $|\psi(0)\rangle$ is prepared as a product state in the computational basis. This is because phase-flip errors (Z-type) lead to a smaller decay in $\langle \avg{Z}(t) \rangle_0$ than bit-flip errors (X- or Y-type)~\cite{S:Ippoliti2021}.
Assuming that phase-flip errors have no effect, i.e., $\rho = I_iZ_j\rho I_iZ_j=Z_iI_j\rho Z_iI_j=Z_iZ_j\rho Z_iZ_j$, the two-qubit depolarising channel in Eq.~(\ref{eq:depolarizingnoise2}) can be modified as
\begin{align}
    \tilde{\mathcal{E}}_{ij}^{(2q)}(\rho)=&\left(1-\frac{4}{5}p_2\right)\rho + \frac{p_2}{15}\sum_{\alpha,\beta}\nolimits' \sigma_{\alpha,i}\sigma_{\beta,j} \rho \sigma_{\alpha,i}\sigma_{\beta,j},
\end{align}
which results in a reduced fidelity $\tilde f :=1-4p_2/5 \sim f^{4/5}$, where $f=1-p_2$.
To correct the underestimated fidelity at $\theta_x=\pi$, we can propose the following modification to the error mitigation formula [Eq.~(4) in the main text]: 
\begin{align}\label{eq:norm-mod}
\langle \avg{Z}(t) \rangle 
\approx
\frac{
\langle \avg{Z}(t) \rangle_{0}}
{{|\langle \avg{Z}(t) \rangle_{0,\theta_x=\pi}|}^{5/4}}.
\end{align}
In this correction, we neglect single-qubit errors, whose probability $p_1$ is smaller than $p_2$.
Figure~\ref{fig:figs8_5over4} shows $\langle \avg{Z}(t) \rangle$ obtained using the modified error mitigation protocol based on Eq.~(\ref{eq:norm-mod}).
Comparing to the mitigation using Eq.~(4) [see Fig.~\ref{fig:L133_peps_en}], the modified protocol yields slightly better agreement with the 2dTNS results, suggesting that accounting for the anisotropy of Pauli errors can improve mitigation accuracy.

\section{Out-of-time-order correlators}\label{sec:otoc}

\begin{figure*}[t]
\includegraphics[width=1.0\textwidth]{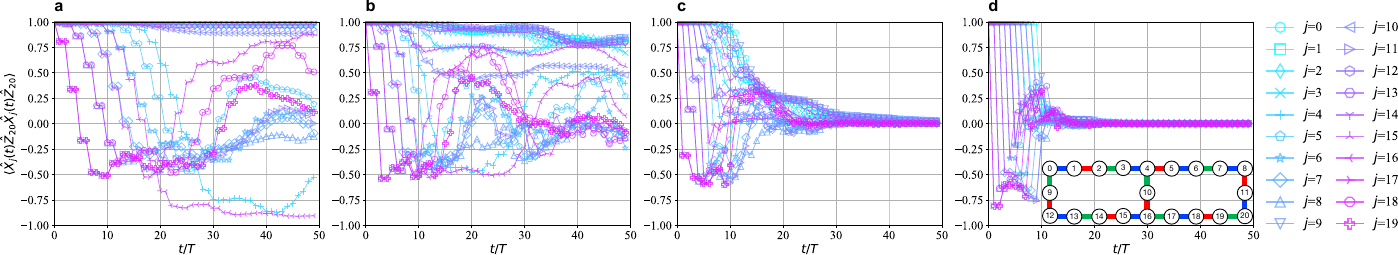}
\caption{
Time evolution of the OTOC $\langle \hat X_j(t) \hat Z_{20} \hat X_j(t) \hat Z_{20} \rangle$ for the kicked Ising model on a heavy-hexagonal lattice of $L=21$ qubits, obtained using the state-vector method. 
        The positions of the qubits $j$ are shown in \cap{d}.
        The parameter $\theta_x$ is \cap{a} $0.9\pi$, \cap{b} $0.8\pi$, \cap{c} $0.7\pi$, and \cap{d} $0.6\pi$, with $\theta_z$ fixed at zero.
        Here, the initial state is prepared as a product state with all qubits set to $|0\rangle$, i.e., a fully-polarised state. 
}
\label{fig:otoc1}
\end{figure*}

\begin{figure*}[t]
\includegraphics[width=1.0\textwidth]{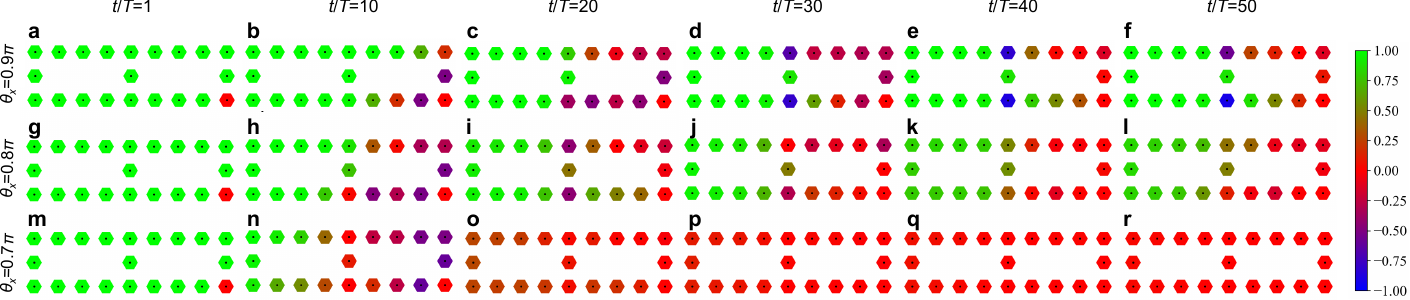}
\caption{
Time evolution of the OTOC $\langle \hat X_j(t) \hat Z_{20} \hat X_j(t) \hat Z_{20} \rangle$ for the kicked Ising model on a heavy-hexagonal lattice of $L=21$ qubits (see the inset of Fig.~\ref{fig:otoc1}d).
The parameter $\theta_x$ is \cap{a}--\cap{f} $0.9\pi$, \cap{g}--\cap{l} $0.8\pi$, and \cap{m}--\cap{r} $0.7\pi$, with $\theta_z$ fixed at zero.
Time steps $t/T$ are; 1 for \cap{a}, \cap{g}, and \cap{m}; 10 for \cap{b}, \cap{h}, and \cap{n}; 20 for \cap{c}, \cap{i}, and \cap{o}; 30 for \cap{d}, \cap{j}, and \cap{p}; 40 for \cap{e}, \cap{k}, and \cap{q}; and 50 for \cap{f}, \cap{l}, and \cap{r}.
Here, the initial state is prepared as a product state with all qubits set to $|0\rangle$, i.e., a fully-polarised state.}
\label{fig:otoc2}
\end{figure*}

In this section, we demonstrate that the DTCs observed in the kicked Ising model on a heavy-hexagonal lattice are realized in a prethermal regime.
To this end, we perform state-vector simulations to evaluate the out-of-time-order correlator (OTOC)~\cite{S:Larkin1969,S:Kitaev2018}, defined as $\langle \hat X_j(t) \hat Z_k \hat X_j(t) \hat Z_k \rangle$, for a system of $L=21$ qubits arranged as shown in the inset of Fig.~\ref{fig:otoc1}d. The results are presented in Figs.~\ref{fig:otoc1} and \ref{fig:otoc2}.
The growth of local operators, i.e., quantum 
scrambling~\cite{S:Hayden2007,S:Sekino2008}, describes the spreading of quantum 
information in time.
OTOCs are widely used as quantitative probes of such scrambling. 
At early times, the OTOC takes a value of 1 becuase $\hat X_j(t)$ and $\hat Z_k$ are initially nonoverlapping and thus commute. 
As scrambling progresses and the operators fail to commute, the OTOC decreases and eventually saturates to 0. 
For $\theta_x=0.9\pi$ and $0.8\pi$, the OTOC remains near 1 or exhibits persistent oscillations, as shown in Figs.~\ref{fig:otoc1}a and \ref{fig:otoc1}b, respectively.
This behavior indicates the absence of significant quantum scrambling up to $t/T<50$.
Since scrambling underpins thermalisation in isolated quantum systems~\cite{S:Deutsch1991,S:Srednicki1994}, its suppression results in the long-lived stability of the prethermal DTCs observed in this regime. 
In contrast, for $\theta_x=0.7\pi$ and $0.6\pi$, the OTOC decays and saturates to 0, as shown in Figs.~\ref{fig:otoc1}c and \ref{fig:otoc1}d, signaling the onset of scrambling. 
The decay is faster at $\theta_x=0.6\pi$ than at $\theta_x=0.7\pi$, where a short-lived prethermal plateau appears around $t/T \sim 20$ (Fig.~\ref{fig:otoc1}c). 
To examine the spatial dependence of scrambing, we show colormaps of $\langle \hat X_j(t) \hat Z_{20} \hat X_j(t) \hat Z_{20} \rangle$ at various time steps in Fig.~\ref{fig:otoc2}.
The operator spreading is clearly slower for $\theta_x=0.9\pi$ and $0.8\pi$ compared to $\theta_x=0.7\pi$.
Indeed, for $\theta_x=0.9\pi$ and $0.8\pi$, several qubits still exhibit OTOC values close to 1 even at $t/T=50$.

\begin{figure*}[t]
\includegraphics[width=1.0\textwidth]{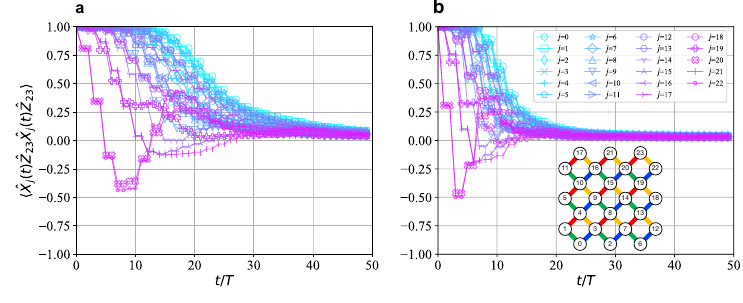}
\caption{
Time evolution of the OTOC $\langle \hat X_j(t) \hat Z_{23} \hat X_j(t) \hat Z_{23} \rangle$ for the kicked Ising model on a square lattice of $L=24$ qubits, obtained using the state-vector method.
        The positions of the qubits $j$ are shown in \cap{b}.
        The parameter $\theta_x$ is \cap{a} $0.9\pi$ and \cap{b} $0.8\pi$, with $\theta_z$ fixed at zero. 
        Here, the initial state is prepared as a product state with all qubits set to $|0\rangle$, i.e., a fully-polarised state. 
}
\label{fig:otoc3}
\end{figure*}

\begin{figure*}[t]
\includegraphics[width=1.0\textwidth]{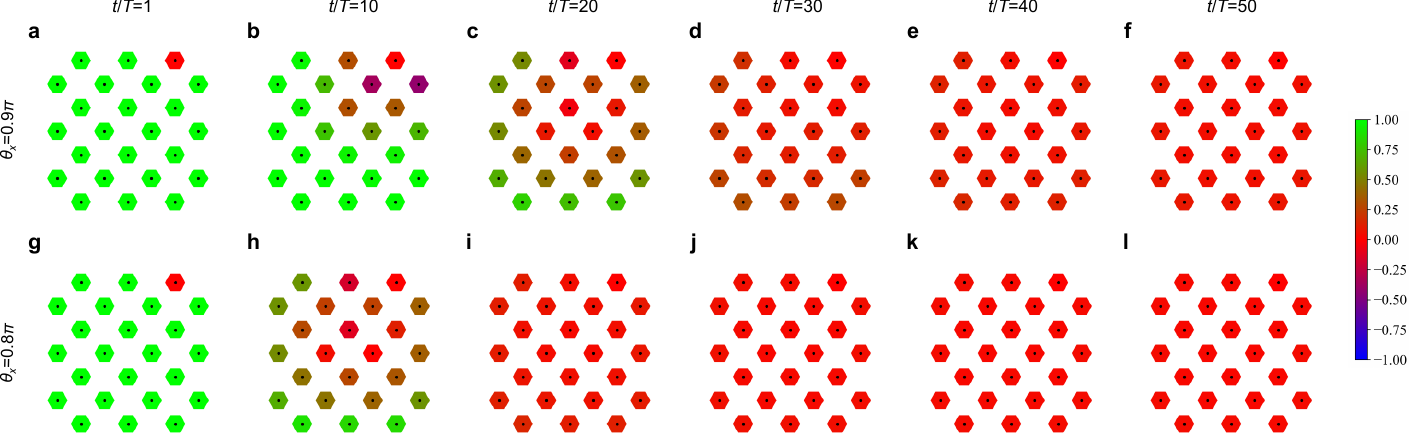}
\caption{
Time evolution of the OTOC $\langle \hat X_j(t) \hat Z_{23} \hat X_j(t) \hat Z_{23} \rangle$ for the kicked Ising model on a square lattice of $L=24$ qubits (see Fig.~\ref{fig:otoc3}b).
The parameter $\theta_x$ is \cap{a}--\cap{f} $0.9\pi$ and \cap{g}--\cap{l} $0.8\pi$, with $\theta_z$ fixed at zero.
Time steps $t/T$ are: 1 for \cap{a} and \cap{g}; 10 for \cap{b} and \cap{h}; 20 for \cap{c} and \cap{i}; 30 for \cap{d} and \cap{j};  40 for \cap{e} and \cap{k}; and 50 for \cap{f} and \cap{l}.
Here, the initial state is prepared as a product state with all qubits set to $|0\rangle$, i.e., a fully-polarised state. 
}
\label{fig:otoc4}
\end{figure*}

For comparison, we also present the OTOC $\langle \hat X_j(t) \hat Z_{23} \hat X_j(t) \hat Z_{23} \rangle$ for the kicked Ising model on a square lattice of $L=24$ qubits, arranged as shown in the inset of Fig.~\ref{fig:otoc3}b.
The results, obtained using the state-vector method, are shown in Figs.~\ref{fig:otoc3} and \ref{fig:otoc4}. 
We find that the OTOCs on the square lattice saturate to zero even for $\theta_x=0.9\pi$, whereas those on the heavy-hexagonal lattice remain far from saturation under the same conditions (see Figs.~\ref{fig:otoc1}a).
This comparison highlights that quantum scrambling proceeds significantly more slowly on the heavy-hexagonal lattice, supporting the emergence of stable prethermal DTCs in this geometry.
Notably, for $\theta_x = 0.9\pi$ and $0.8\pi$, we observe a clear suppression of quantum-information propagation at qubits $j=4$ and $16$. 
This behaviour arises because symmetry charges accumulate at these coordination-3 qubits, effectively acting as blocking points for information flow (see Ref.~\cite{S:Shinjo2025} for details).  
In contrast, at $\theta_x = 0.7\pi$, the entanglement propagation overcomes these blocking sites, allowing information to spread throughout the lattice and leading to full quantum scrambling.

\section{Long-time dynamics} \label{sec:longtime} 

\begin{figure*}[t]
\includegraphics[width=1.0\textwidth]{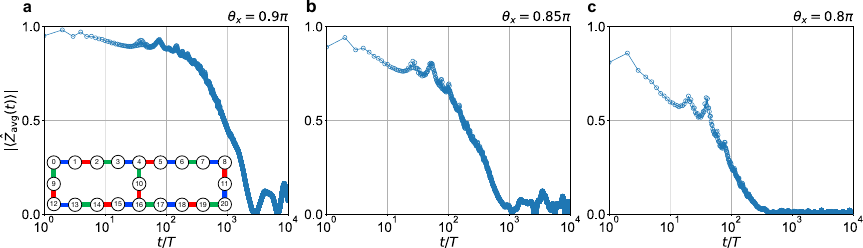}
\caption{
    Long-time evolution of the magnetisation $|\langle \avg{Z}(t)  \rangle|$ for the kicked Ising model on a heavy-hexagonal lattice of $L=21$ qubits (see the inset in \cap{a}), obtained using the state-vector method. 
    The parameter $\theta_x$ is \cap{a} $0.9\pi$, \cap{b} $0.85\pi$, and \cap{c} $0.8\pi$, with $\theta_z$ fixed at zero.
    The average is taken over the entire systems, with the initial state prepared as a product state with all qubits set to $|0\rangle$, i.e., a fully-polarised state. }
\label{fig:longtime21}
\end{figure*}

\begin{figure*}[t]
\includegraphics[width=1.0\textwidth]{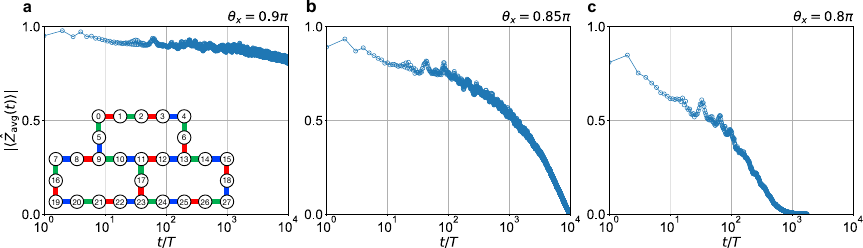}
\caption{
    Same as Fig.~\ref{fig:longtime21}, but for a heavy-hexagonal lattice of $L=28$ qubits (see the inset in \cap{a}).}
\label{fig:longtime28}
\end{figure*}

In this section, we present the long-time dynamics of magnetisation  oscillations in the kicked Ising model on a heavy-hexagonal lattice.
It is well known that Floquet systems with a small transverse field perturbation $\theta_x \sim \pi$ can exhibit two-step relaxation dynamics~\cite{S:Mori2016, S:Else2017}.
Following an initial rapid relaxation, the system enters a metastable prethermal state characterized by slow energy absorption. Eventually, it relaxes to an infinite-temperature state, as predicted by the Floquet eigenstate thermalization hypothesis (ETH)~\cite{S:Dalessio2013, S:Ponte2015}.

Figures~\ref{fig:longtime21} and \ref{fig:longtime28} show the absolute value of the magnetisation, $|\langle \hat Z_\text{avg}(t) \rangle|$, averaged over the entire system, for heavy-hexagonal lattices with $L=21$ qubits (see the inset of Fig.~\ref{fig:longtime21}a) and $L=28$ qubits (see the inset of Fig.~\ref{fig:longtime28}a), respectively. These results are  obtained using the state-vector method with the initial state $|\psi(0)\rangle$ prepared as a product state with all qubits seto to $|0\rangle$.
For the $L=21$ qubit system, prethermal plateau-like structures are observed in the time intervals $10 < t/T < 200$, $10 < t/T < 70$, and $10 < t/T < 30$ for $\theta_x = 0.9\pi$, $0.85\pi$, and $0.8\pi$, respectively. 
These plateau structures persist for longer durations in the $L=28$ qubit system, appearing in the intervals $10 < t/T < 1000$, $10 < t/T < 200$, and $10 < t/T < 70$ for the same values of $\theta_x$.
This system size dependence suggests that prethermal DTCs on the heavy-hexagonal lattice become increasingly robust with larger system sizes.

Together with the behavior of the OTOCs discussed in Sec.~\ref{sec:otoc}, these results support the conclusion that prethermal DTCs emerge in the parameter range $0.8\pi \alt \theta_{x} < \pi$ on the heavy-hexagonal lattice.


\section{Tensor network simulations: 2dTNS } \label{sec:peps}

\begin{figure}[htbp]
\includegraphics[width=0.5\hsize]{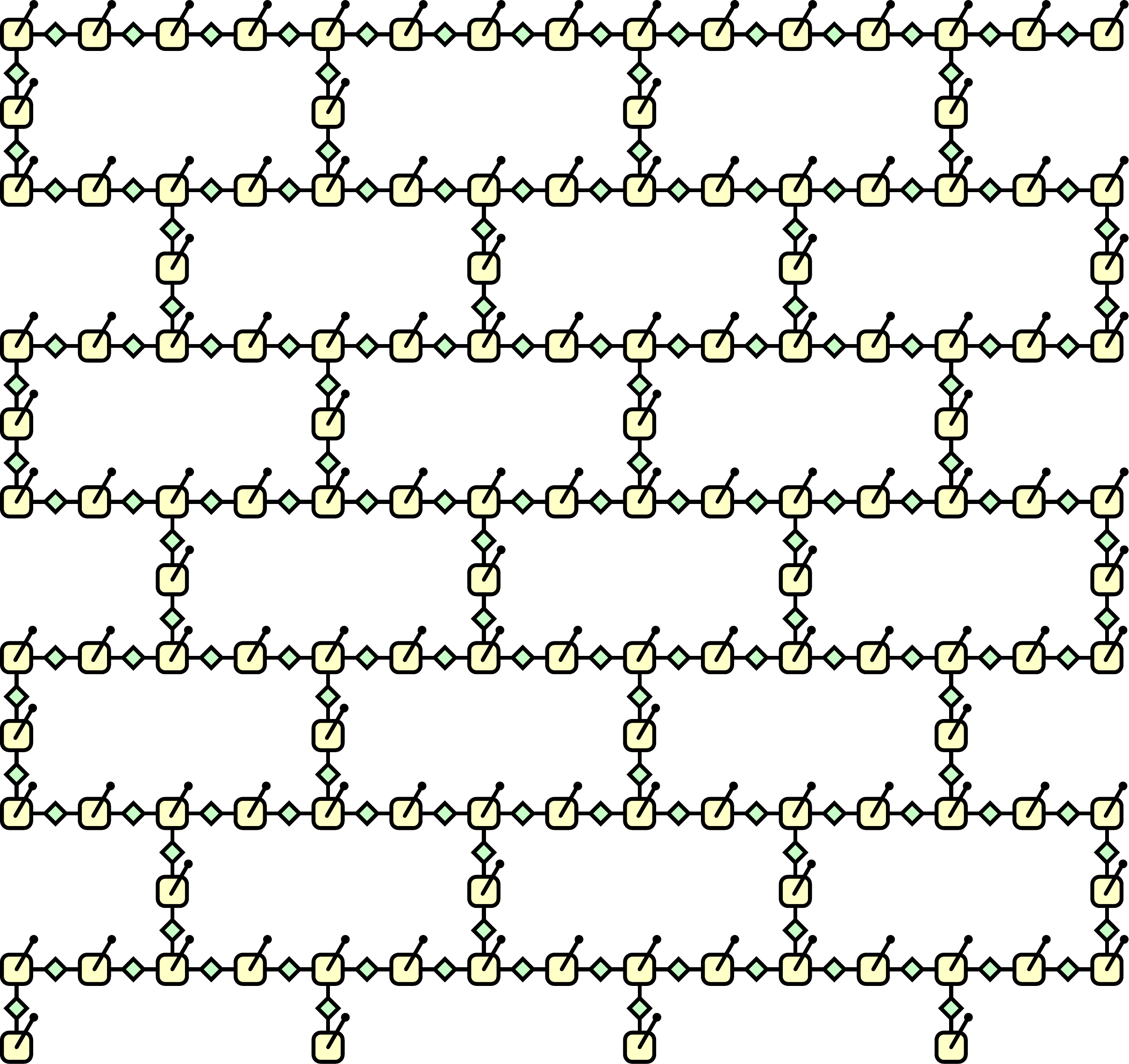}
\caption{
Schematic structure of the two-dimensional gauging tensor network for the heavy-hexagonal lattice comprising $L=133$ qubits. 
Edges represent virtual bonds with bond dimension $\chi$.
Vertex tensors, denoted as squares, have virtual bonds along with a physical bond representing qubit degrees of freedom.
Gauge tensors, depicted as diamonds, are positioned on the edges connecting neighboring vertex tensors.
}
\label{fig:gtns}
\end{figure}

In this section, we describe our 2dTNS method based on a gauging tensor network technique. 
Gauging tensor network states (TNSs) provide a compact representation of quantum states within an approximation framework~\cite{S:Vidal03,S:Vidal04,S:Shi06,S:Vidal07,S:Orus08,S:Nagaj08,S:Kalis12,S:Ran12,S:Phien15,S:Tindall23a,S:Tindall23b}. 
The gauging tensor network consists of two types of tensors: vertex tensors and gauge tensors. 
The vertex tensor has a physical bond representing qubit degrees of freedom, as well as virtual bonds that reflect the topology of the system. 
The gauge tensor is positioned on the edge connecting neighboring vertex tensors. 
The structure of the gauging tensor network for the IBM Quantum Heron processor, {\torino}, which forms the heavy-hexagonal lattice, is illustrated in Fig.~\ref{fig:gtns}. 
For the heavy-hexagonal lattice, there are two types of vertex tensors: one with two virtual bonds and the other with three virtual bonds. 
The dimension of the virtual bond, i.e., bond dimension $\chi$, controls the accuracy of the approximation. 
In our study, $\chi$ denotes the maximum bond dimension utilised in a TNS.

\begin{figure}[htbp]
  \includegraphics[width=0.5\hsize]{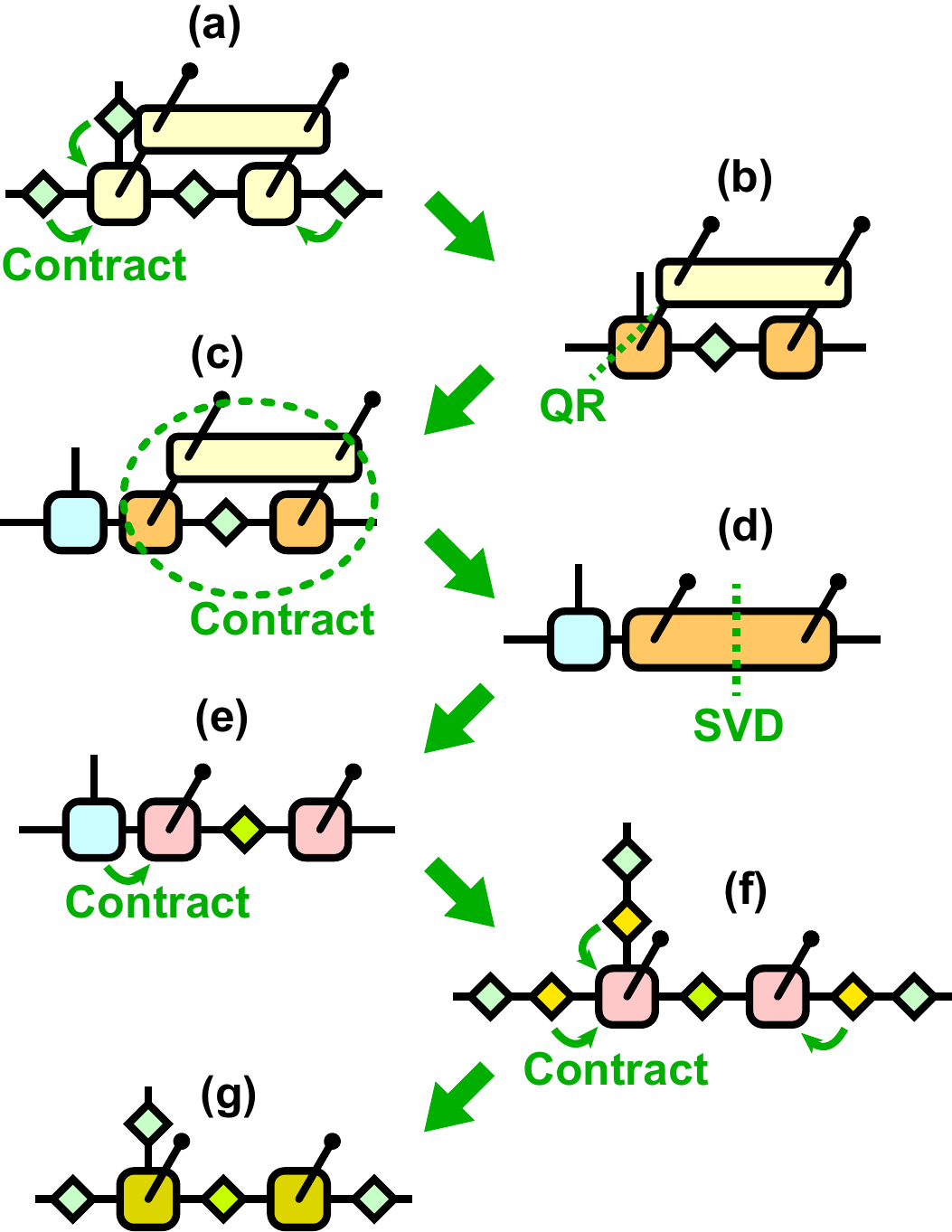}
  \caption{A depiction of the time-evolved block decimation (TEBD) process for a TNS given in Fig.~\ref{fig:gtns}.
    (a) The surrounding gauge tensors are absorbed into the vertex tensors to which a two-qubit gate is applied.
    (b) QR decomposition is performed on the vertex tensor which has the three virtual bonds to reduce the computational cost for the following singular value decomposition (SVD) 
    procedure.
    (c) All tensors connected to the two-qubit gate and the gauge tensor between the two vertex tensors are contracted.
    (d) SVD is performed on the tensor obtained in step (c).
    (e) The isometric tensor obtained from the QR decomposition in step (b) is incorporated into the new tensor obtained by SVD in step (d). 
    (f) Inverses of the surrounding gauge tensors are computed and 
    absorbed into the vertex tensors which have the physical bonds.
    (g) New tensors are obtained. 
  }
  \label{fig:tebd}
\end{figure}

To apply a two-qubit gate to a gauging TNS with bond dimension $\chi$, we employ the time-evolving block decimation (TEBD) method~\cite{S:Vidal03}, outlined in Fig.~\ref{fig:tebd}.
Initially, in step (a), the surrounding gauge tensors and the vertex tensors targeted for the two-qubit gate are contracted. Following this, 
QR decomposition is performed on the vertex tensor in step (b) to optimise the computational efficiency for the subsequent singular value decomposition (SVD) in step (d). 
All tensors connected to the two-qubit gate, along with the gauge tensor, are contracted in step (c). 
SVD is then executed on the resulting tensor in step (d), yielding a total of $2\chi$ singular values. 
To prevent an increase in bond dimension, only the $\chi$ largest singular values are retained, with the remaining space truncated to dimension $\chi$. 
Subsequently, this truncated diagonal matrix, composed of the $\chi$ largest singular values, replaces the original gauge tensor, with the neighboring vertex tensors updated, in steps (e)-(g).

\begin{figure}[htbp]
  \includegraphics[width=0.45\hsize]{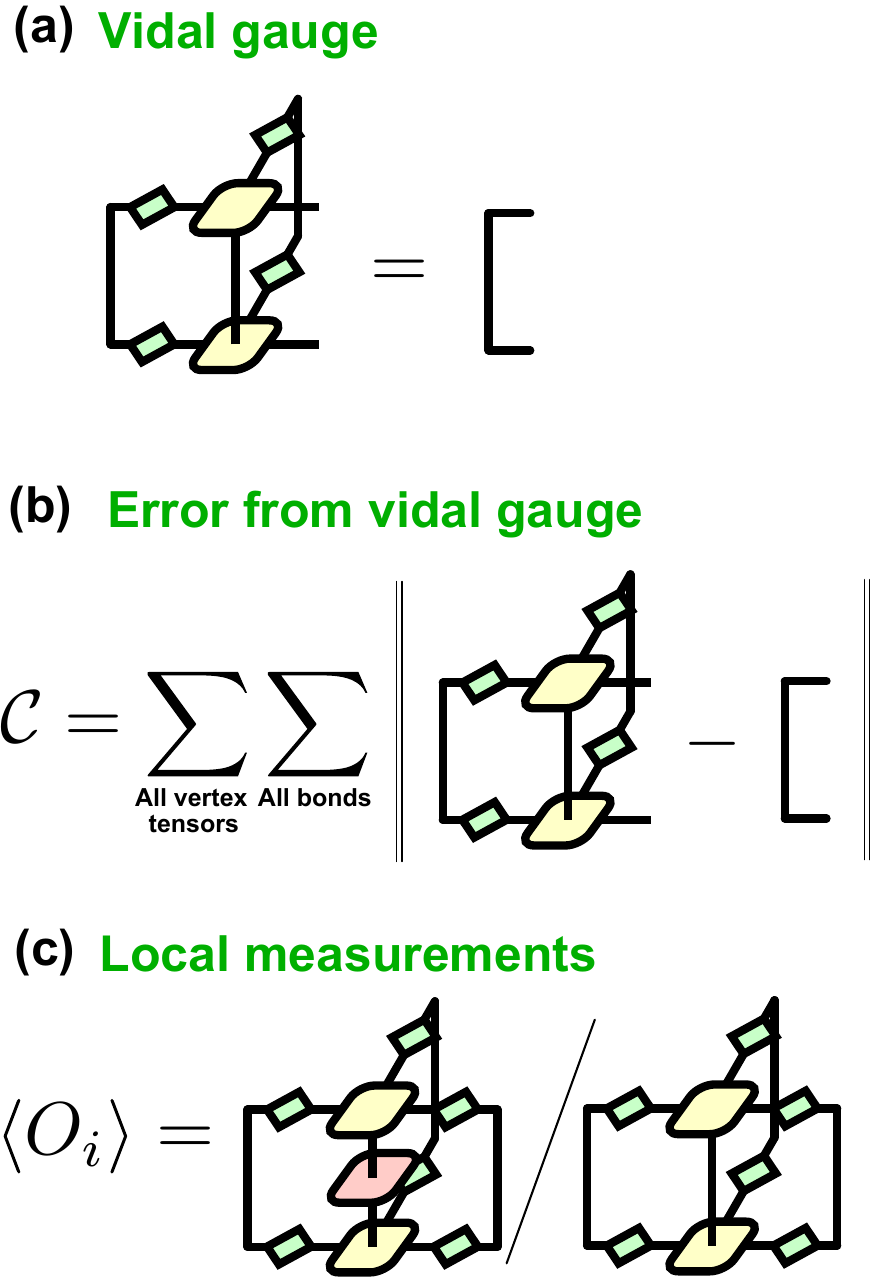}
  \caption{(a) A TNS in the Vidal gauge satisfies the isometric condition after absorbing all but one of their neighboring gauge tensors around the vertex tensors.
    (b) Deviation from the Vidal gauge is quantified with $\mathcal{C}$. 
    Here, $\vvert \mathcal{A} \vvert$ represents the Frobenius norm for a tensor 
    $\mathcal{A}$. 
    Each tensor $\mathcal{A}$ after contraction is assumed to be normalised, i.e., $\tilde{\mathcal{A}} = \mathcal{A} / \vvert \mathcal{A} \vvert$. 
    (c) Tensor contraction process for evaluating the expectation value $\langle \hat{O}_i \rangle$ of a local operator $\hat{O}_i$ (denoted as a red square), assuming that 
    a TNS is in the Vidal gauge.
  }
  \label{fig:vidalgauge}
\end{figure}

More generally, in a gauging TNS, gauge tensors are typically represented as diagonal matrices. 
While a gauging TNS can describe the same quantum state,
there exist various approaches to implement the gauge at an edge.
One particularly notable gauge is the canonical gauge, also known as Vidal's gauge.
Within this framework of the Vidal gauge, the vertex tensors become isometric after absorbing all but one of the gauge tensors on their adjacent bonds,
as schematically shown in Fig.~\ref{fig:vidalgauge}(a). 
A notable advantage of employing the Vidal gauge is evident 
when computing the expectation value of a local physical quantity.
For instance, when evaluating the expectation value of a physical quantity at a single site, the laborious task of contracting the entire 
tensor network is replaced with local tensor contraction, as shown in Fig.~\ref{fig:vidalgauge}(c).
This presents a significant advantage, particularly in the context 
of two-dimensional TNSs,
where full contraction becomes computationally infeasible without resorting to 
approximate techniques.

\begin{figure}[htbp]
  \includegraphics[width=0.5\hsize]{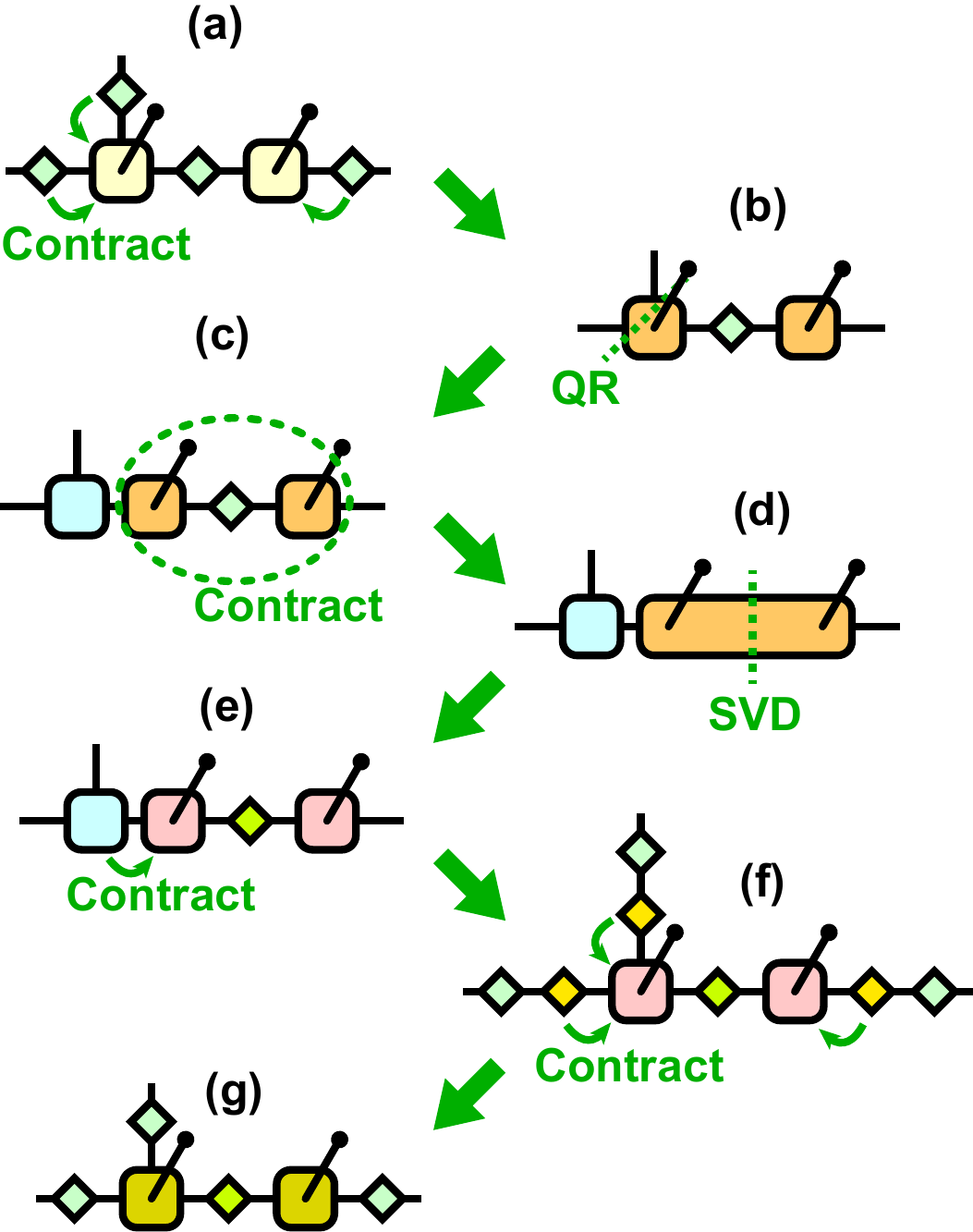}
  \caption{Procedure for the trivial simple update (tSU),
    essentially equivalent to the process outlined in Fig~\ref{fig:tebd}, but without 
    the inclusion of the two-qubit gate.}
  \label{fig:surg}
\end{figure}

However, the TEBD process generally disrupts the Vidal gauge,
even when we start with a TNS with the Vidal gauge. 
A procedure to restore this broken gauge is known as regauging. 
Particularly, simple and well-known regauging methods include trivial simple update (tSU)~\cite{S:Jiang08,S:Corboz10,S:Jahromi19} and belief propagation~\cite{S:Tindall23a,S:Tindall23b,S:Leifer08,S:Poulin08,S:Robeva18,S:Alkabetz21}. 
In our study, we adopt the tSU regauging method outlined in Fig.~\ref{fig:surg}. 
The tSU is a procedure essentially equivalent to that described in Fig.~\ref{fig:tebd}, except that the two-qubit gate is replaced with identity. 
Repeatedly applying this procedure across all edges in the tensor network improves the gauge and ultimately restores the Vidal gauge, in principle. 
While there is a mathematical proof for this in a tree tensor network~\cite{S:Alkabetz21},
it is not established for a general tensor network with a loop structure. 
In our case, we assess the error $\mathcal{C}$ from the Vidal gauge, as defined in Fig.~\ref{fig:vidalgauge}(b), at each time step,
and iterate the tSU until it sufficiently converges. 
Moreover, we observe no significant difference between taking the Vidal gauge at every time step and only before measuring a physical quantity. 
This suggests that the accuracy of a TNS itself at each time step is not greatly influenced by whether the Vidal gauge is taken or not. 
Additionally, we performed calculations using the belief propagation~\cite{S:Tindall23a,S:Tindall23b}, and found that the results coincide with those obtained using the tSU,
indicating convergence to the same fixed point~\cite{S:Alkabetz21}.

We utilise this method, referred to as the 2dTNS method, to simulate the quantum dynamics governed by the Floquet operator $\hat{U}_{\rm F}$ described in the main text.
Note that the single-qubit unitary gates $\hat{R}_{Z_i}(\theta_z)$ and $\hat{R}_{X_i}(\theta_x)$ can be treated exactly within this framework.  
As benchmark calculations, in Fig.~\ref{fig:L28_peps}, we compare the results of $\langle \avg{Z}(t) \rangle$ for the heavy-hexagonal lattice of $L=28$ qubits obtained by the 2dTNS method with various bond dimensions ($\chi=40$, 100, 300, and 400) with those obtained 
by the state-vector method. 



\begin{figure}[htbp]
  \includegraphics[width=0.5\hsize]{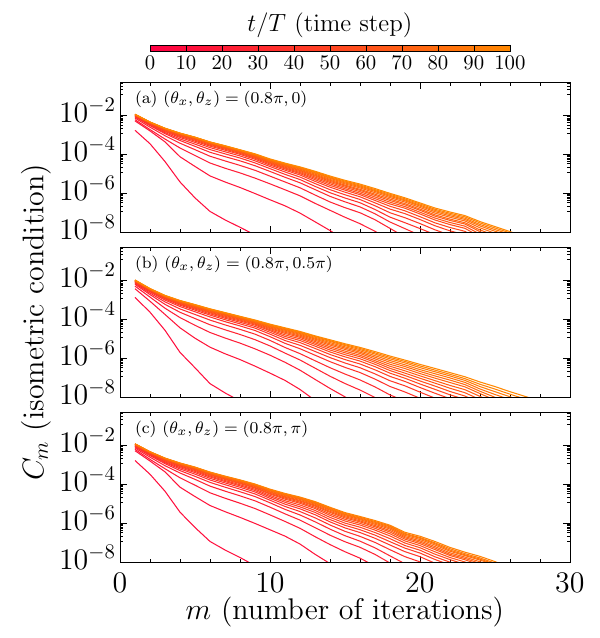}
  \caption{
  Convergence to the Vidal gauge vs. the number $m$ of iterations when employing the tSU regauging for the heavy-hexagonal lattice comprising $L=133$ qubits. 
    The parameters $(\theta_x,\theta_z)$ are (a) $(0.8\pi,0)$, (b) $(0.8\pi, 0.5\pi)$, and (c) $(0.8\pi,\pi)$. The bond dimension of the TNS is $\chi=200$.
    $C_m$ is the error from the Vidal gauge, defined as in Fig.~\ref{fig:vidalgauge}(b), at iteration $m$, 
    with color intensity denoting the time step $t/T$ (indicated at the top).
    We set the error tolerance of $10^{-8}$ to ensure compliance with the Vidal gauge in our simulations. 
   }
  \label{fig:ciso}
\end{figure}

Finally, Fig.~\ref{fig:ciso} shows typical results for the convergence of a TNS towards that with the Vidal gauge when excuting the tSU regauging iterations during the simulations for the heavy-hexagonal lattice consisting of $L=133$ qubits, which corresponds to the results presented in Figs.~\ref{fig:L133_peps} and \ref{fig:L133_peps_en}. 
In this figure, each iteration of the tSU regauging process covers all the edges across the entire lattice. 
As the time step $t/T$ progresses, the number of iterations required for achieving a fixed accuracy also increases. 
Nevertheless, even at $t/T=100$, the convergence remains excellent with fewer than 30 
iterations. 


